\documentclass{article}
\usepackage{amsmath,amsthm,amsfonts,latexsym}

\newcommand{\Reali}{{\mathbb{R}}}
\newcommand{\Complessi}{{\mathbb{C}}}
\newcommand{\Naturali}{{\mathbb{N}}}
\newcommand{\Interi}{{\mathbb{Z}}}

\newcommand{\cC}{{\cal C}}
\newcommand{\cA}{{\cal A}}

\newcommand{\cH}{{\cal H}}

\newcommand{\cI}{{\cal I}}
\newcommand{\cN}{{\cal N}}

\newcommand{\cM}{{\cal M}}

\newcommand{\cJ}{{\cal J}}
\newcommand{\cR}{{\cal R}}

\newcommand{\cX}{{\cal X}}
\newcommand{\cK}{{\cal K}}

\newcommand{\cZ}{{\cal Z}}

\newcommand{\PSL}{{\mbox{\tt PSL}(2,\Reali)}}
\newcommand{\Pnot}{{\mbox{\tt P}_0}}
\newcommand{\liePnot}{{{\mathfrak p}_0}}
\newcommand{\liePSL}{{{\mathfrak s}{\mathfrak l}(2,\Reali)}}
\newcommand{\covPSL}{{\overline{\mbox{\tt PSL}}(2,\Reali)}}

\newcommand{\Mob}{{\mbox{\tt Mob}}}
\newcommand{\cinfcerchio}{{{\cal C}^{\infty}(S^1,\Reali)}}

\newcommand{\cinfcom}{{\cal C}^{\infty}_{0}}
\newcommand{\fphi}{\hat{\phi}}
\newtheorem{Thm}{Theorem}[section]
\newtheorem{Cor}[Thm]{Corollary}
\newtheorem{Prop}[Thm]{Proposition}
\newtheorem{Lemma}[Thm]{Lemma}
\theoremstyle{definition}
\newtheorem{Dfn}{Definition}

\theoremstyle{remark}
\newtheorem{rem}{Remark} 
 
\title{\huge  Extensions of Conformal Nets\\
 and Superselection Structures}
\author{
D. Guido$^{(1)}$, R. Longo$^{(1)}$, H.-W. Wiesbrock$^{(2)}$\\
{}\\
$^{(1)}$Dipartimento di Matematica,\\ Universit\`a di Roma ``Tor
Vergata'',\\ Via della Ricerca Scientifica, I--00133 Roma, Italy.\\
E-mail: guido@mat.utovrm.it, longo@mat.utovrm.it\\
{}\\
$^{(2)}$Freie Universit\"at Berlin,\\
Institut f\"ur Theoretische Physik,\\
Arnimallee 14,
D-14195 Berlin, Germany.\\
E-mail : wiesbroc@physik.fu-berlin.de}
\footnotetext[1]{Supported in
part by Ministero della Pubblica Istruzione and CNR-GNAFA.}
\footnotetext[2]{Supported by the DFG, SFB 288 ``Differentialgeometrie
und Quantenphysik''}

\date{June 26, 1997}
\begin{document}
\maketitle
\markboth{D. Guido, R. Longo and H.W. Wiesbrock}{Extensions of conformal nets}
\renewcommand{\sectionmark}[1]{}

\begin{abstract}
Starting with a conformal Quantum Field Theory on the real line, we show  that
the dual net is still conformal with respect  to a new representation of
the M\"obius group. We infer from this that every  conformal net is normal and
conormal, namely the local von Neumann  algebra associated with an interval
coincides with its double  relative commutant inside the local von Neumann
algebra associated  with any larger interval. The net and the dual net give
together rise   to an infinite dimensional symmetry group, of which we study a
class  of positive energy irreducible representations.
We mention how superselsection sectors extend to the dual net and we
illustrate  by examples how, in general, this
process generates  solitonic sectors.
We describe the free theories associated with the lowest weight $n$
representations of $\PSL$, showing that they violate 3-regularity for $n>2$.
When $n\geq2$, we obtain examples of non M\"obius-covariant sectors of a
3-regular (non 4-regular) net.\end{abstract}

\newpage
\setcounter{section}{-1}
\section{Introduction.}\label{sec:intro}

Haag duality is one of the most important properties in Quantum Field
Theory for the analysis of the superselection structure. It basically
says that the locality principle holds maximally.

Concerning Quantum Field Theory on the usual Minkowski spacetime,
duality may be always assumed, in a Wightman theory, because wedge
duality automatically holds and one can enlarge the net to a dual net
without affecting the superselection structure \cite{[Bi-Wi],[Ro]}.

Nevertheless there might be good reasons
in lower dimensional theories for Haag duality not to be
satisfied, \cite{[Mue]}.
An important case occurs in Conformal QFT: as such a theory
naturally lives on a larger spacetime, duality may fail on the
original spacetime because contributions at infinity are possibly not
detectable there. Moreover in low spacetime dimensions, the
superselection structure of the dual net may change due to the occurrence
of soliton sectors \cite{[St],[Ro]}, and a new information being contained
in the
inclusion of the two nets.

This paper is devoted to an analysis of conformal QFT on the real
line, namely one-dimensional components of a two-dimensional chiral conformal
QFT.

The first aspect that we discuss concerns the symmetries of the dual
net. Starting with a conformal net $\cA$ on $\Reali$, the
Bisognano-Wichmann property holds automatically true \cite{[BGL],
[Ga-Fr]}, thus
the dual net
$$\cA^{d}(a,b)\doteq \cA(-\infty,b)\cap\cA(a,\infty)\ \ \ \ a<b$$
on $\Reali$ is local and obviously
translation-dilation covariant with respect to the same
translation-dilation unitary representation. We shall however prove
that $\cA^{d}$ is even conformally covariant with respect to a new
unitary representation of $\PSL$.

The construction of the new
symmetries is achieved by a new characterization of local conformal
precosheaves on the circle in terms of what we call a +hsm (half-sided
modular) factorization, namely a quadruple $(\cN_i,i\in\mathbb Z_3;\Omega)$
where the $\cN_i$'s are mutually commuting von Neumann algebras with a
joint cyclic separating vector $\Omega$ such that $\cN_i\subset\cN_{i+1}$
is a +hsm modular inclusion  in the sense of
 \cite{[Wi2]} for all $i\in\mathbb Z_3$. This
characterization makes only use of the modular operators and not of
the modular conjugations as in \cite{[Wi3]}.

As second result we shall deduce a structural property for any
conformal net $\cA$: $\cA$ is automatically normal and conormal, namely
if $I\subset\tilde I$ is an inclusion of proper intervals then
\begin{gather}
  \cA(I)^{cc}=\cA(I),\\
  \cA(\tilde I)=\cA(I)\vee \cA(I)^{c}
\end{gather}
where $\cX^{c}=\cX'\cap\cA(\tilde I)$ denotes the relative
commutant in $\cA(\tilde I)$  and $\cX^{cc} =(\cX^{c})^{c}$.
This property is useful in the analysis of the superselection
structure, as discussed below.

Next issue will be to compare the net $\cA$ with the dual net
$\cA^{d}$. We shall give a detailed study of the inclusion
$\cA\subset\cA^{d}$ in a particular model, namely when $\cA$ is the net
associated
with the $n$-th derivative of the $U(1)$-current
algebra, the latter turning out to coincide with $\cA^{d}$. This is a
first quantization analysis and to this end we shall give formulas
relating two irreducible lowest weight representations of $\PSL$ that
agree on the upper triangular matrix subgroup $\Pnot$.
In other words we are studying a class of representations of a
certain infinite dimensional Lie group, the amalgamated free product
$\PSL*_{\Pnot}\PSL$, where a classification is
simply obtained.
 For example an
explicit formula for the unitary $\gamma$ whose second quantization
 implements the canonical
endomorphism associated with the inclusion
$\cA(-1,1)\subset\cA^{d}(-1,1)$ (the product of the ray inversion
unitaries of the two nets) will be given as
a function of the skew-adjoint generator $E$ of the
dilation one-parameter subgroup
$$
\gamma =\frac{(E-1)(E-2)\cdots (E-n)}{(E+1)(E+2)\cdots (E+n)}.
$$
We shall show that the net $\cA$ is not $4$-regular if $n\geq 2$ and
not $3$-regular if $n\geq 3$, where $\cA$ is said to be $k$-regular
if  $\cA$ remains irreducible after removing $k-1$ points from
$\Reali$. The $4$-regularity property played a role in the
covariance analysis in \cite{[Gu-Lo]} where the problem of its general
validity remained open.

We generalize the construction of the Buchholz, Mack and Todorov sectors
for the current algebra (\cite{[BMT]}) to the $n$-th derivative of the
current algebra,
showing that they form a group isomorphic to $\Reali^{2n+1}$, and that
none of them is covariant w.r.t. the conformal group (if $n\neq0$). In
particular this shows results in \cite{[Gu-Lo]} to be
optimal, at least on the real line.

Finally we shall illustrate by examples how sectors of $\cA$ localized in a
bounded interval may have extension to $\cA^{d}$ with soliton
localization.

\section{Structural properties of conformal local precosheaves on
  $S^1$.}\label{sec:first}

In the algebraic approach, see \cite{[Haa]}, chiral
conformal field theories are described as conformally covariant
local precosheaves $\cA$ of von
Neumann algebras on proper intervals of the circle $S^1$.
We start by reviewing some aspects of this framework.

An open  interval $I$ of $S^{1}$ is called {\it proper} if $I$ and the
interior $I'$ of its complement are not empty. The circle will be
explicitly described either as
the points with modulus one in $\Complessi$ or as the one-point
compactification of $\Reali$, these two description being related by the
Cayley transform:
$C:S^1\to\Reali\cup\{\infty\}$ given by
 $z\to i(z+1)(z-1)^{-1}.$
 The group $\PSL$ acts on $S^1$ via its action on $\Reali\cup\{\infty\}$
as fractional transformations. Intervals are labeled either by the
coordinates on $\Reali$ or by complex coordinates of the
 endpoints in $S^{1}\subset\Complessi,$ where in the later case
intervals are represented in positive cyclic order.

A precosheaf $\cA$ is a covariant functor from the
category $\cJ$ of proper intervals with inclusions as arrows to the category
of von Neumann algebras on a Hilbert space $\cH$ with inclusions as arrows,
i.e., a map
$$
I\to\cA(I)
$$
that satisfies:
\begin{itemize}
\item[{\bf A.}] {\bf Isotony}. If $I_{1}\subset I_{2}$ are proper
intervals, then
\begin{equation*}
  \cA(I_{1})\subset\cA(I_{2}).
\end{equation*}
\end{itemize}
The precosheaf $\cA$ will be a (local) {\it conformal precosheaf} if in
addition it satisfies the following properties:
\begin{itemize}
\item[{\bf B.}] {\bf Locality}. If $I_{1}$ and $I_{2}$ are disjoint proper
intervals,
then
\begin{equation*}
  \cA(I_{1})\subset\cA(I_{2})',
\end{equation*}
\item[{\bf C.}] {\bf Conformal invariance}.
There exists a strongly
continuous unitary representation $U$ of $\PSL$ on $\cH$ such that
\begin{equation*}
  U(g)\cA(I) U(g)^*\  =\ \cA(gI),\qquad g\in\PSL,\ I\in\cJ.
\end{equation*}
\end{itemize}
\begin{itemize}
\item[{\bf D.}] {\bf Positivity of the energy}. The generator of the rotation
subgroup $U(R(\cdot))$ (conformal Hamiltonian) is positive.
Here $R(\vartheta)$ denotes the rotation
of angle $\vartheta$ on $S^1$.
\item[{\bf E.}] {\bf Existence of the vacuum}. There exists a unit vector
$\Omega\in\cH$ (vacuum vector) which is  $U(\PSL)$-invariant and cyclic for
$\vee_{I\in\cJ}\cA(I)$.
\end{itemize}

The Reeh-Schlieder Theorem now states that the vacuum vector $\Omega$
is cyclic and separating for any local algebra $\cA(I).$

Let us recall that uniqueness of the vacuum is equivalent both to the
irreducibility
of the precosheaf or to the factoriality property for local algebras.
We shall denote by $\Mob$ the M\"obius group, namely the group of
conformal transformations in $\Complessi$ that leave the unit circle
globally invariant. The group $\PSL$ is then identified with the subgroup of
orientation preserving transformations and $\Mob$ is generated by
$\PSL$ and an involution.

Let $I_1$ be the upper semi-circle  parameterized as
$(0,+\infty)$; we associate to $I_1$ the following two
one-parameter subgroups of $\Mob :$
First the dilations (relative to $I_1$),
\begin{equation*}
 \Lambda_{I_1}(t)=\begin{pmatrix} e^{-t/2} &  0\\ 0&
 e^{t/2}\end{pmatrix},
\end{equation*}
leaving $I_1$ globally stable, second the translations
\begin{equation*}
  T_{I_1}(s)=\begin{pmatrix} 1& s\\ 0& 1 \end{pmatrix} .
\end{equation*}
mapping $I_1$ into itself for positive $s\geq 0.$
In general, if $I$ is any interval in $S^1$, there exists a
$g\in\PSL$ such that $I=gI_1$ and we set
\begin{equation*}
\Lambda_I =g\Lambda_{I_1}g^{-1},\qquad T_I =g T_{I_1} g^{-1}.
\end{equation*}
The definition of the dilations does not depend on $g$, while
the translations of $I$ are defined up to a rescaling of the parameter,
that however
does not play any role in the following, because we are only
interested in the subgroups generated by them.

The subgroup generated by $\Lambda_{I_1}(\cdot)$ and  $T_{I_1}(\cdot)$,
denoted  by $\Pnot$, is the subgroup of upper triangular matrices of
$\PSL$ and plays an important role in the following, especially in the next
Section.

We shall associate with any proper interval $I$ a diffeomorphism $r_I$
of $S^1$,
the reflection mapping $I$ onto the causal complement  $I^\prime$,
i.e. fixing the boundary
points of $I$. In the  case of
 $I_1 = (0,+\infty)$
\begin{equation*}
r_{I_1} x = -x
\end{equation*}
and one can extend this definition to a generic $I$ as before.
Notice that $r_{I_1}$ is orientation reversing.

By a (anti-)representation $U$ of $\Mob$ we shall mean the obvious
generalization of the notion of unitary representation where
$U(r_{I})$ is anti-unitary.

For a general conformal precosheaf the {\it Bisognano-Wichmann Property} holds
\cite{[BGL],[Ga-Fr]}:
$U$ extends to a  unitary (anti-)representation of $\Mob$
such that, for any $I\in\cJ$,
\begin{align}
U(\Lambda_I (2\pi t))&=\Delta_{I}^{it},\label{BW}\\
U(r_I) &= J_I,
\end{align}
where $\Delta_I$ and $J_I$ are the Tomita-Takesaki modular operator and
modular conjugation associated with $(\cA(I),\Omega).$
This implies Haag duality:
$$
\cA(I)'=\cA(I'),\quad I\in\cI.
$$
Let now $(\cN\subset\cM,\Omega)$ be a triple where $\cN\subset\cM$ is an
  inclusion of von Neumann algebras acting on a Hilbert space
  $\cH$ and $\Omega\in\cH$ is a cyclic and separating vector for $\cN$ and
  $\cM$.
\begin{enumerate}
\item  $(\cN\subset\cM,\Omega)$ is said to be {\sl standard}  if $\Omega$ is
cyclic also for the relative commutant  $\cN^c\doteq \cM\cap\cN'$ of $\cN$  in
$\cM$, see \cite{[Do-Lo]}.
\item If $\sigma_{t}^{\cM}$ denotes the modular automorphism associated to
$(\cM,\Omega)$, then the  triple $(\cN\subset\cM, \Omega)$ is said to be
{\sl $\pm$
half-sided modular} ($\pm$ hsm) if $\sigma_{-t}^{\cM}(\cN) \subset\cN$ for,
respectively, all $t\ge 0$ or all $t\le 0$.
\item A {\sl $\pm$hsm  factorization} of von Neumann algebras is a quadruple
$(\cN_0,\cN_1,\cN_2,\Omega)$, where $\{ \cN_i, i\in \mathbb Z_3\}$
is a set of pairwise commuting von Neumann algebras, $\Omega$ is a
cyclic separating vector for each $\cN_i$ and
$(\cN_i\subset\cN_{i+1}',\Omega)$ is a $\pm$hsm inclusion for each
$i\in\mathbb Z_3$.
\end{enumerate}
In the work \cite{[Wi4]}, local conformal precosheaves have been
characterized in terms of $\pm$hsm standard inclusions of von Neumann
algebras and the adjoint action of the modular
conjugations. (This work is based on a statement about hsm modular
inclusions \cite{[Wi1]}, whose correct proof is contained in
\cite{[ArZs1]}.)
  We shall give here below an
alternative characterization in
terms of a $\pm$hsm factorization, that has the advantage of using only
the modular groups and not the modular conjugations.

\begin{Lemma}\label{abstractPSL}
 Let $G$ be the universal group (algebraically) generated by
3 one-parameter subgroups
$\Lambda_i(\cdot),\ i\in \mathbb Z_3$, such that $\Lambda_i$ and
$\Lambda_{i+1}$ have
the same commutation relations of $\Lambda_{I_i}$ and $\Lambda_{I_{i+1}}$ for
each $i\in \mathbb Z_3$, where $I_0,I_1,I_2$ are intervals forming a
partition of $S^1$. Then $G$ is isomorphic to $\covPSL$, the universal
covering group of $\PSL$, and the $\Lambda_i$'s are continuous one
parameter subgroups naturally corresponding to $\Lambda_{I_i}$.
\end{Lemma}
\begin{proof} Obviously $G$ has a quotient isomorphic to $\covPSL$,
and we denote by $q$ the quotient map. As the exponential map is
a local diffeomorphism of the  Lie algebra of a Lie group and the Lie
group itself, there exists a neighbourhood $\cal U$ of the origin
$\mathbb R^3$ such that the map $(t_0,t_1,t_2)\to
\Lambda_{I_0}(2\pi t_0)\Lambda_{I_{1}}(2\pi t_1)\Lambda_{I_{2}}(2\pi t_2)$
is a diffeomorphism of $\cal U$ with a neighbourhood of the identity of
$\covPSL$. Therefore the map   $\Phi: (t_0,t_1,t_2)\in{\cal{U}}\to
\Lambda_{0}(2\pi t_0)\Lambda_{1}(2\pi t_1)\Lambda_{2}(2\pi t_2)\in G$ is
still one-to-one.
It is
easily checked that
the maps $g\Phi: {\cal {U}}\to G$, $g\in G$, form an atlas on $G$, thus $G$ is a
manifold. In fact $G$ is a Lie group since the group operations are
smooth, as they are locally smooth. Now $G$ is connected by
construction and $q$ is a local diffeomorphism of $G$ with $\covPSL$,
hence a covering map, that has to be an isomorphism because of the
universality property of $\covPSL$.
\end{proof}
\begin{Thm}\label{char}
  Let $(\cN_0,\cN_1,\cN_2,\Omega)$ be a {\sl +hsm  factorization} of
  von Neumann algebras and let $I_0,I_1,I_2$ be intervals forming a
  partition of $S^1$ in counter-clockwise order.  There exists a unique local
   conformal precosheaf $\cA$ on $S^{1}$
   such that $\cA(I_i)=\cN_i$, $i\in\mathbb Z_3$, with $\Omega$ the
   vacuum vector. The (unique)
   positive energy unitary representation
   $U$of $\PSL$ is determined by the modular prescription
   $U(\Lambda_{I_i}(2\pi t))=\Delta_{I_i}^{it}$.
\end{Thm}
Notice that every +hsm factorization of von Neumann algebras arises by
considering the von Neumann algebras associated to 3 intervals of $S^1$
as in the above theorem, due to the geometric property of the modular
group (\ref{BW}).
\begin{proof} The subgroup of $\PSL$ generated by the one-parameter
subgroups $\Lambda_{I_i}(2\pi t)$ and $\Lambda_{I_{i+1}}(2\pi s)$,
 $i\in\mathbb Z_3$, is a two-dimensional Lie group $\mbox{\tt P}_i$
isomorphic to the translation-dilation group $\mbox{\tt P}_0$ .
As $(\cN_i,\cN_{i+1}',\Omega)$ is a +hsm standard inclusion, by a
result first stated in \cite{[Wi1]} with an erroneous proof and whose
correct proof is given in \cite{[ArZs1]}, the unitary group generated by
$\Delta_{I_i}^{it}$
and $\Delta_{I_{i+1}}^{is}$ is isomorphic to $\mbox{\tt P}_i$, indeed
there exists a unitary representation $U_i$ of $\mbox{\tt P}_i$
determined by $U_i(\Lambda_{I_i}(2\pi t))=\Delta_{I_i}^{it}$
and $U_i(\Lambda_{I_{i+1}}(-2\pi s))=\Delta_{I_{i+1}}^{is}$, therefore
by Lemma \ref{abstractPSL}, there exists a unitary representation $U$ of
 $\covPSL$, such
that $U|_{\mbox{\tt P}_i}=U_i$.

Let $t_0=\frac{1}{2\pi} \ln 2.$ Then we have (\cite{[Wi1], [ArZs1]},
see the remarks above)
\begin{equation}\label{last}
\mbox{Ad }\Delta_{I_{0}}^{it_0}\Delta_{I_{1}}^{it_0} (\cN_0)=\cN_1',
\end{equation}
and similarly
\begin{equation}\label{pi}
\mbox{Ad }\Delta_{I_{2}}^{it_0}\Delta_{I_{0}}^{it_0}
\Delta_{I_{1}}^{it_0}\Delta_{I_{2}}^{it_0}
\Delta_{I_{0}}^{it_0}\Delta_{I_{1}}^{it_0}(\cN_0 ) = \cN_0'.
\end{equation}
The element $\Lambda_{I_{2}}(2\pi t_0)\Lambda_{I_{0}}(2\pi t_0)
\Lambda_{I_1}(2\pi t_0)\Lambda_{I_{2}}(2\pi t_0)
\Lambda_{I_{0}}(2\pi t_0)\Lambda_{I_{1}}(2\pi t_0)$ is easily seen to
be conjugate to the rotation by $\pi$
in $\PSL$, hence equation (\ref{pi}) entails $U(2\pi)$ to implement an
automorphism on  $\cN_0$.

Set $\cA(I_0):= \cN_0.$ If $I$ is an interval of $S^1$, then
$I=gI_{0}$ for some
$g\in\PSL$,
and we set $\cA(I)=U(g)\cA(I_0)U(g)^*$.

Since  the group $G_{I_0}$ of all $g\in\covPSL$
such that $gI_0=I_0$ is generated by $\Lambda_{I_0}(t),\ t\in
\mathbb R$ and
by rotations of $2k\pi, \ k\in\mathbb Z$, then
$U(g)\cA(I_0)U(g)^*=\cA(I_0)$ for all $g\in G_{I_0}$ and the von Neumann
algebra
$\cA(I)$ is well defined.

The isotony of $\cA$ follows if we show that
$gI_0\subset I_0\implies \cA(gI_0)\subset \cA(I_0)$. Indeed  any such
$g$ is a product of an element in $G_{I_0}$ and translations
$T_{I_0}(\cdot)$ and  $T_{I'_0}(\cdot)$ mapping $I_0$ into itself,
hence the isotony follows by the half-sided modular conditions.

By (\ref{last}) we have
$$
\mbox{Ad }
\Delta_{I_{1}}^{it_0}\Delta_{I_{2}}^{it_0}
\Delta_{I_{0}}^{it_0}\Delta_{I_{1}}^{it_0}(\cN_0 ) = \cN_2
$$
and since the corresponding element in $\covPSL$ maps $I_0$ onto
$I_2$, we get $\cN_2=\cA(I_2)$ and analogously $\cN_1=\cA(I_1)$.

The locality of $\cA$ now follows by the factorization property.

Finally $U$ is a true representation of $\PSL$ by the vacuum conformal
spin-statistics theorem \cite{[GuLo2]}, and the positivity of the energy
follows by the
Bisognano-Wichmann property (\ref{BW}), see \cite{[Wi3],[Wi4]}.
\end{proof}
Although a conformal precosheaf satisfies
Haag duality on $S^1$, duality on $\Reali$ does not necessarily
hold.
\begin{Lemma}\label{stradd}
Let $\cA$ be a local conformal precosheaf on $S^{1}$. The following are
equivalent:
\begin{itemize}
\item [$(i)$]The restriction of $\cA$ to $\Reali$ satisfies Haag duality:
$$
\cA(I)=\cA(\Reali\backslash I)'
$$
\item [$(ii)$] $\cA$ is {\rm{strongly additive}}: If $I_1$, $I_2$ are the
connected components of the interval $I$ with one internal point removed, then
$$
\cA(I)=\cA(I_1)\vee\cA(I_2)
$$
\item [$(iii)$] if $I, I_1, I_2$ are intervals as above
$$
\cA(I_1)'\cap\cA(I)=\cA(I_2)
$$
\end{itemize}
\end{Lemma}
\begin{proof} Note that by M\"obius covariance we may suppose that the
point removed in $(i)$ and $(ii)$ is the point $\infty$. Now
 $(i)\Leftrightarrow(ii)$ because $\Reali\backslash I$
consists of two contiguous intervals in $S^{1}$ whose union has closure
equal $I'$, and by Haag duality $\cA(I)=\cA(I')'$. Similarly
$(ii)\Leftrightarrow(iii)$ because, after taking commutants and
renaming the intervals, one relation
becomes equivalent to the other one.
\end{proof}
Examples of conformal precosheaf on $S^1$ that are not strongly
additive, i.e. not Haag dual on the line, were first given in
 \cite{[Hi-Lo],[Bu-Sch]}
and \cite{[Y]}. We will look in some detail at an example of \cite{[Y]}
in Section~3.
Haag duality on $S^1$ entails duality for half-lines on $\Reali$ hence
essential duality, namely the dual net of the restriction $\cA_{0}$
to $\Reali$ is local:
\begin{equation*}
  I\mapsto \cA_{0}^d(I)\doteq \cA(\Reali\backslash I)',\qquad I\subset\Reali.
\end{equation*}
Due to locality the net $\cA_{0}^d$ is larger than the original one, namely
\begin{equation*}
  \cA_{0}(I)\subset \cA_0^d(I),\qquad I\subset\Reali.
\end{equation*}
$\cA_{0}^d$ is usually called the dual net,
see \cite{[Bi-Wi],[Bu-Sch]} and its main
 feature is that it obeys Haag duality on $\Reali$.
The  dual net does not in general transform covariantly under the covariance
representation of the starting net.
\begin{Thm}\label{BW=conformal}
  Let $\cA$ be a local net of von Neumann algebras on the
  intervals of
  $\Reali$, $\Omega$ a cyclic and separating vector for the von
  Neumann algebra $\cA(I)$ associated with each interval $I\subset\Reali$
  and $U$ a $\Omega$-fixing unitary representation of the
  translation-dilation group acting covariantly on $\cA$. The
  following are equivalent:
  \begin{itemize}
  \item [$(i)$] $\cA$
  extends to a conformal precosheaf on $S^{1}$.
  \item [$(ii)$] The Bisognano-Wichmann property holds for $\cA$, namely
\begin{equation}\label{BWR}
 \Delta_{\Reali^{+}}^{it}=U(\Lambda_{\Reali^{+}}(2\pi t)).
 \end{equation}
  \end{itemize}
  \end{Thm}
\begin{proof}$(i)\Rightarrow (ii)$: See \cite{[BGL],[Ga-Fr]}.

$(ii)\Rightarrow (i)$:
Note first that, by translation covariance,
$\Delta_{(a,\infty)}^{it}=U(\Lambda_{(a,\infty)}(2\pi t)) $
for all $a\in\mathbb R$. Hence $\cA(-\infty, a)$ is a
von Neumann subalgebra of $\cA(a,\infty)'$ that is cyclic on $\Omega$
and globally invariant under the modular group of $\cA(a,\infty)'$
with respect to $\Omega$, hence, by the Tomita-Takesaki theory,
duality for half-lines holds
$$
\cA(a,\infty)'=\cA(-\infty, a).
$$
Recall now that if $\cN\subset\cM$ is an inclusion of von Neumann algebras
and $\Omega$ is a cyclic and separating vector for both $\cN$ and
$\cM$, then $(\cN\subset\cM,\Omega)$ is +hsm iff
$(\cM'\subset\cN',\Omega)$ is $-$hsm \cite{[Wi1], [ArZs1]}.

Then it is immediate to check
$(\cA(-\infty,-1),\cA(-1,1),\cA(1,\infty),\Omega)$ to be a +hsm
factorization
of von Neumann algebras, so we get a conformal precosheaf from
Theorem \ref{char}. Due to Bisognano-Wichmann property this is indeed an
extension to $S_1$ of the original net.
\end{proof}
Note as a consequence that a local net on $\mathbb R$ as above with
property (\ref{BWR}) automatically has a PCT symmetry, namely
$$ J_{\Reali^{+}}\cA(I)J_{\Reali^{+}}=\cA(-I),\quad \forall \:
  \text{interval}\: I\subset\Reali .$$
Now, if $\cA$ is a local conformal precosheaf on $S^1$, its restriction
$\cA_{0}$
to $\Reali$ does not depend, up to isomorphism, on the point we cut
$S^1$, because of M\"obius covariance. The local precosheaf on $S^1$
extending $\cA^d_0$ is thus well defined up to isomorphism. We call it the
dual precosheaf of $\cA$ and denote it by $\cA^{d}$.

\begin{Cor}\label{confdual}
  The dual precosheaf of a conformal precosheaf on $S^1$ is a strongly additive
  conformal precosheaf on $S^1$.
\end{Cor}
\begin{proof} By construction, the dual net satisfies Haag duality on $\Reali$,
hence strong additivity by Lemma~\ref{stradd}.
\end{proof}
\begin{rem}
Let us compare the precosheaves $\cA$ and $\cA^d$ on $S^1$. First we
observe that equality holds if and only if the conformal precosheaf $\cA$
is strongly additive.
As mentioned above locality implies
$$\cA(I) \subset \cA^d (I) \ \ \ \mbox{if } -1 \notin I,
$$
i.e. if $I$ does not contain the point infinity, while Haag duality on
$S^1$ implies
$$
\cA^d (I) \subset \cA(I) \ \ \ \mbox{ if } -1 \in I.
$$
 Therefore the observable algebras assciated with
bounded intervals of the real line are enlarged while the others, associated
with intervals
containing the point at infinity, decrease.
The observable algebras associated with half-lines, i.e. with intervals having
$-1$ as
a boundary point, remain fixed.
Due to the Bisognano-Wichmann property, which holds for all conformal
precosheaves, altering the algebras implies a change in the
representation of the conformal group $\PSL.$ Moreover, since the
algebras associated with half-lines coincide, both representations agree on the
isotropy group of the point at infinity, i.e. on the subgroup
$\mbox{\tt{P}}_0$ of $\PSL$
generated by translations and dilations.
\end{rem}

An inclusion $\cN\subset\cM$ of von~Neumann algebras is said to be
{\it normal} if $\cN\ =\ \cN^{cc}$, where $\cX^{c}=\cX'\cap \cM$ denotes the
relative commutant of $\cX$ in $\cM$, and {\sl conormal} if $\cM$ is generated
by $\cN$ and its relative commutant  w.r.t. $\cM$, i.e., $\cM\ =\ \cN\vee\cN^c$
(i.e. $\cM'\subset\cN'$ is normal).

We shall then say that a local conformal precosheaf $\cA$ is (co-)normal if
the inclusion $\cA(I_1)\subset \cA(I_2)$ is (co-)normal for any pair
$I_{1}\subset I_{2}$ of proper intervals of  $S^{1}$. By Haag duality,
normality  and conormality are equivalent properties of conformal
precosheaves.
\begin{Thm}\label{confnorm}
  Any conformal precosheaf on $S^1$ is normal and conormal.
\end{Thm}
\begin{proof}
  Let us consider first an inclusion of two proper intervals
  $I_{1}\subset I_{2}$ with a common boundary point.

If $\cA$ is strongly additive, the inclusion of von Neumann algebras
$\cA(I_{1})\subset \cA(I_{2})$ is conormal as in this case
$\cA(I_{1})'\cap\cA(I_{2})=\cA(I_{2}\backslash I_{1})$.
In the general case, by conformal invariance we may assume that $I_{1}$
and $I_{2}$ are respectively
  the intervals of the real line $(1,+\infty)$ and $(0,+\infty)$. By
  definition then $\cA(I_{1})=\cA^{d}(I_{1})$,
  $\cA(I_{2})=\cA^{d}(I_{2})$,
  with $\cA^{d}$ the dual net, hence the inclusion $\cA(I_{1})\subset
\cA(I_{2})$
  is conormal by  Corollary ~\ref{confdual} and the above argument.
  As  $\cA(I_{2})'\subset\cA(I_{1})'$ is conormal,  $\cA(I_{1})\subset
\cA(I_{2})$
  is also normal.

  It remains to show the normality of $\cA(I_{1})\subset\cA(I_{2})$ when
$I_{1}\subset I_{2}$ are intervals with no common boundary point, e.g.
$I_1=(b,c)$ and $I_2=(a,d)$, with $a<b<c<d$. Then we set $I_3=(a,c)$ and
$I_4=(b,d)$, therefore $I_{1}=I_{3}\cap I_{4}$ and both $I_3$ and $I_4$
are subintervals of $I_2$ with a common boundary point. Then
the double relative commutant of $\cA(I_{1})$ in $\cA(I_{2})$ is given by
\begin{equation}
\cA(I_{1})^{cc}\subset\cA(I_{3})^{cc}\cap\cA(I_{4})^{cc}
=\cA(I_{3})\cap\cA(I_{4})=\cA(I_{1})
\end{equation}
where the last equality is a consequence of duality and additivity and
implies the first inclusion; the opposite inclusion is elementary
\end{proof}
\begin{Cor}\label{normal}
  Let $(\cN\subset \cM,\Omega)$ be a +hsm  standard inclusion of von
  Neumann algebras. In this case:
   \begin{itemize}
  \item The inclusion $\cN\subset\cM$ is normal and conormal.
  \item There exists a unique strongly additive local conformal precosheaf $\cA$
  of von Neumann algebras on $S^1$
  with $\cM=\cA(0,+\infty)$, $\cN=\cA(1,+\infty)$, and
  $\Omega$ the vacuum vector.
  \item There exists a bijection between local conformal precosheaves $\cA$
  of von Neumann algebras on $S^1$
  with $\cM=\cA(0,+\infty)$, $\cN=\cA(1,+\infty)$,
  $\Omega$ the vacuum vector, and  von Neumann subalgebras $\cN_0$ of
  $\cN'\cap\cM$ cyclic on $\Omega$ such that
  $(\cN_0\subset\cM,\Omega)$ is a $-$hsm inclusion and $(\cN_0 \subset
  \cN',\Omega)$ is a $+$hsm inclusion.
  \end{itemize}
\end{Cor}
\begin{proof}
Starting with the last point, notice that $(\cM',\cN_0,\cN,\Omega)$
is a +hsm
factorization of von Neumann algebras, and clearly any
 +hsm
factorization arises in this way, therefore the thesis is a
consequence of Theorem \ref{char}.

In the special case $\cN_0=\cN'\cap\cM$ we then obtain the second
statement by Lemma \ref{stradd}(ii)$\Longleftrightarrow$ (iii).

The first point is then a consequence of Theorem \ref{confnorm}.
\end{proof}
Note  as a consequence that a hsm modular standard inclusion
$(\cN\subset\cM,\Omega)$ is
also {\it pseudonormal} :
$$
\cN\vee J\cN J = \cM\cap J\cM J
$$
where $J$ is the modular conjugation of $(\cN'\cap \cM,\Omega)$ and
has the continuous interpolation property (see \cite{[Do-Lo]}).

The irreducibility of $\cA$ in Corollary~\ref{normal}  is equivalent in
particular to the factoriality of $\cN$  and $\cM$, see \cite{[GuLo2]}. This is
also equivalent to the center of $\cN$ and $\cM$ to  have trivial intersection.
Thus we have the following.
\begin{Cor}  Let $(\cN\subset \cM,\Omega)$ be +hsm and standard. Then
$\cN$ and $\cM$ have the same center $\cZ$ and
$(\cN\subset \cM,\Omega)$  has a direct integral decomposition
$\cN=\int_{\cZ}^{\oplus}\cN_{\lambda}\text{d}\mu(\lambda)$,
$\cM=\int_{\cZ}^{\oplus}\cM_{\lambda}\text{d}\mu(\lambda)$, $\Omega =
\int_{\cZ}^{\oplus}\Omega_{\lambda}\text{d}\mu(\lambda)$, where each
$(\cN_{\lambda}\subset\cM_{\lambda},\Omega_{\lambda})$ is either a  +hsm
standard inclusion of $III_{1}$ factors or trivial ($\cN=\cM=\mathbb C$).
\end{Cor}
\begin{proof}
The modular group acts trivially on the center, so that
$$
\mbox{Ad } \Delta_{\cM}^{it} ( \cN \cap \cM') = \cN \cap \cM', \ \ \
\forall t \in \Reali .
$$
Since $\mbox{Ad }\Delta_\cM^{it_0}\cN=\cM$ for a suitable $t_0$, we
immediately obtain
$\cN \cap\cM' = \cM \cap \cM'$ and $$
\cM \cap \cM' = \cN \cap \cM' \subset \cN \cap \cN',
$$
i.e. $\cZ ( \cM) \subset \cZ (\cN).$ Using the commutants we obtain the
equality and the direct integral decomposition as stated. Applying
\cite{[Wi1]}, Theorem 12, and \cite{[ArZs1]}, we finish the proof.

\end{proof}
The following  Corollary summarizes part of the above discussion,
based on results in \cite{[Bo],[Wi1], [ArZs1]}.
\begin{Cor}
There exists a one-to-one correspondence between:
\begin{itemize}
\item Isomorphism classes of standard +half-sided modular
inclusions $(\cN\subset\cM,\Omega)$
\item Isomorphism classes of Borchers triples $(\cM,U,\Omega)$, $($i.e.
$\cM$ is a von Neu\-mann algebra with a cyclic separating unit vector
$\Omega$ and $U$ is a one-parameter $\Omega$-fixing unitary group
with positive generator s. t. $U(t)\cM U(-t)$ $\subset$ $\cM$, $t>0)$
such that $\Omega$ is cyclic for $U(t)\cM' U(-t)\cap\cM$ for some,
hence for all, $t>0$.
\item Isomorphism classes of translation-dilation covariant,
Haag dual nets on $\Reali$ with the Bisognano-Wichmann property
$\Delta_{\Reali^+}^{it}=U(\Lambda(2\pi t))$.
\item Isomorphism classes of strongly additive
local conformal precosheaves of von
Neumann algebras on $S^{1}$.
\end{itemize}
\end{Cor}
The notion of isomorphism in the above setting has an obvious meaning.
Note however that an isomorphism between local conformal precosheaves
can be defined as an isomorphism of precosheaves relating the
vacuum states, as in this case it will automatically intertwine the
M\"obius representations as these are unique, being fixed by the
modular prescriptions.

\section{Representations of $\PSL$ and derivatives of the $U(1)$-current}
\label{sec:second}
\subsection{A class of representations of $\PSL*_{\Pnot} \PSL$}
\label{subsec:classpsl2r}
In section~\ref{sec:first} we have seen that we may associate with any
conformal precosheaf on
$S^1$ another conformal precosheaf on $S^1$ which is its Haag dual net on
$\Reali$. This
amounts to ``cut the circle,'' namely to fix a special point (``$\infty$'')
and to redefine the local algebras associated to intervals which are
relatively compact in $S^1\setminus\{\infty\}$ in such a way that Haag duality
holds on $S^1\setminus\{\infty\}$. The representations of
$\PSL$ associated with the two nets
coincide when restricted to the group $\Pnot$ generated by
translations and dilations, therefore give a representation of
$\PSL*_{\Pnot}\PSL$, i.e., the free product of two copies of $\PSL$
amalgamated by the subgroup $\Pnot$.

Let us denote by $i_1$, resp. $i_2$, the embeddings of $\PSL$ into
the first, resp. the second, component of the free product, and by $i$ the
immersion of $\Pnot$ in the amalgamated free product.
Then we shall consider  on $\PSL*_{\Pnot}\PSL$
the  topology generated by the maps $i_1$ $i_2$, namely a unitary
 representation $U$ of $\PSL*_{\Pnot}\PSL$ is strongly continuous if and only if
$U\circ i_1$  and $U\circ i_2$ are strongly continuous.

We shall classify the class of strongly continuous unitary positive
energy  irreducible representations of
$\PSL*_{\Pnot}\PSL$ whose restrictions $U\circ i_k$ are still irreducible or,
equivalently, such that $U((H_1 - H_2 )T)$ is
a scalar, where $T$ is the generator of the translations belonging to
$\liePnot$, the Lie algebra of $\Pnot$ and $H_{k}= i_k (H)$ are the
generators of
the rotation subgroup. This amounts to classify
the  unitary positive energy representation with
$U((H_1 - H_2 )T)$ central, as these decompose into a direct integral of
irreducible representations in the previous class.  As we shall see,
this is the general case in a free theory.

\begin{Thm}\label{freeprod}
Let $U$ be an irreducible unitary representation of $\PSL*_{\Pnot}\PSL$ with
positive energy, namely $-iU(T)$ is positive.
Then $U\circ i_k$ is irreducible for some $k=1,2$ if and only if both the
$U\circ
i_k$ are irreducible, and if and only if $U((i_1 (H)-i_2
(H))i(T))\in\Complessi$, where $H$
resp. $T$ generate rotations  resp. translations in $\PSL$.
Moreover, such representations are classified by pairs of
natural numbers $(n_1 ,n_2 )$, where  $n_k$ is the lowest weight of the
representation $U\circ i_k$ of $\PSL$.
\end{Thm}
As is known the matrices
\begin{equation*}
  E=\frac{1}{2}\begin{pmatrix} 1&  0\\ 0& -1\end{pmatrix}, \qquad
  T=\begin{pmatrix} 0& 1\\ 0& 0 \end{pmatrix}, \qquad
  S=\begin{pmatrix} 0& 0\\  -1& 0\end{pmatrix}.
\end{equation*}
form a basis for the Lie algebra $\liePSL$
and verify the commutation relations
\begin{equation}\label{CommRel}
[E,T]=T,\quad [E,S]=-S,\quad [T,S]=-2E\ .
\end{equation}
Let us remind us that the conformal Hamiltonian is
$H=\frac{i}{2}(T+S)$ and that the lowest weight of the representation
is its lowest eigenvalue.
The Casimir operator
\begin{equation}\label{Casimir}
\lambda = E(E-1) - TS
\end{equation}
is a central element of the universal enveloping Lie algebra, thus its value
in an irreducible unitary representation is a scalar.
If $U$ is a unitary irreducible  non-trivial lowest weight
representation of $\PSL$, then the selfadjoint  generator $-iU(T)$ of
$U(e^{tT})$
is positive and non-singular, therefore $U(e^{tE})$ and
$e^{it\log(-iU(T))}$ give a
representation  of the Weyl commutation relations, namely $U$ restricts to a
unitary representation of $\Pnot$, that has to be irreducible because any
bounded operator commuting with $E$ and $T$ also commutes with $S$ due to the
formula (\ref{Casimir}).
 The von Neumann uniqueness  theorem then implies that
the restriction of $U$ to $\Pnot$ is unitarily equivalent to the Schr\"odinger
representation, therefore $E\mapsto d/dx$, $T\mapsto -ie^{x}$ on $L^2(\Reali)$.

We now describe all lowest weight representations of $\PSL$ (or its
universal covering group $\covPSL$)
as extensions of the representation of $\Pnot$.

Let us fix now the unitary irreducible representation of $\PSL$ with lowest
weight 1 and denote by $E_0$, $T_0$ and $S_0$ the image in this representation
of the above Lie algebra generators $E$, $T$ and $S$.

\begin{Prop}\label{Prop:restrictionpsl2r}

\begin{itemize}
\item
Each non-trivial irreducible  unitary representation $U$ of
$\covPSL$ with lowest weight $\geq 1$  is unitarily equivalent to the
representation
 obtained by
exponentiation of the operators $T_\lambda =T_0$, $E_\lambda =E_0$,
$S_\lambda=S_0 -\lambda T_0 ^{-1}$, for some $\lambda>0$\footnotemark.
  \item
  All $\lambda>0$ appear and $\lambda=\alpha(\alpha -1)$ if
   $U$ has lowest weight
  $\alpha$
  \item
   $\lambda$
  may be written as $\lambda=\frac{m}{n}(\frac{m}{n} -1)$, $m,n\in\Naturali$,
  if and only if $U$ is a representation of the n-th covering of $\PSL$
  \end{itemize}
  \footnotetext{If $A$ and $B$ are linear operator with closable sum,
   the closure of their sum is denoted simply by $A+B$.}
\end{Prop}
\begin{proof}
The first two statements follow from the above discussion since the value of
the Casimir operator in the unitary representation with lowest weight
$\alpha$ is
equal to $\lambda=\alpha(\alpha -1)$, see \cite{[L]}, and one gets the
formula for $S_\lambda$ by
multiplying both sides of (\ref{Casimir}) by $T^{-1}$.

To check the last point, first observe that when $\lambda\ge 0$,
$\lambda=\alpha(\alpha -1)$,  $\alpha\ge 1$, we get an orthonormal set of
eigenvectors for the (self-adjoint) conformal Hamiltonian
  \begin{equation*}
    H_{\lambda}=\frac12\left(e^x -\frac{d}{dx}\left(e^{-x}\frac{d}{dx}\right)
    +\lambda e^{-x}\right).
  \end{equation*}
In fact, set $\phi_{\alpha}=e^{\alpha x} e^{-e^{x}}$ and define the following
operators $a_{\pm}^{\alpha}= 2 E_\lambda\pm i(T_\lambda +S_\lambda )$. We also
set for simplicity of notation $H\doteq  H_{\alpha(\alpha-1)}$. Since
$Ha^{\alpha}_{\pm}=a_{\pm}^{\alpha}(H\pm 1)$,
$H\phi_{\alpha}=\alpha\phi_{\alpha}$ and  $a_{-}^{\alpha}\phi_{\alpha}=0$ then
$\phi_{\alpha}^n\doteq   (a_{+}^{\alpha})^n \phi_{\alpha}$ is an orthogonal set
of eigenvectors of $H$ with eigenvalues $\alpha +n$. An application of the
Stone-Weierstrass theorem shows that it is actually a  basis, and the generated
vector space is a G{\aa}rding domain for $H_\lambda$, $T$, $E$. The rest of the
statement follows easily.
\end{proof}
{\bf Proof of Theorem~\ref{freeprod}:}
If $U((i_1 (H)-i_2 (H))i(T))$ is a scalar, $U\circ i_2(e^{tH})$ belongs to
$(U\circ i_1(\PSL*_{\Pnot}\PSL))''$ and $U\circ i_1(e^{tH})$ belongs to
$(U\circ i_2(\PSL*_{\Pnot}\PSL))''$, therefore, since $U$ is irreducible,
$U\circ i_k$ is irreducible too, $k=1,2$. On the other hand, if say $U\circ
i_1$ is
irreducible, we may identify it with one of the representations described in
 Proposition~\ref{Prop:restrictionpsl2r} for some $\alpha\in\Reali$.
Then, since
$U$ is irreducible and $U\circ i_1|_\Pnot=U\circ i_2|_\Pnot$, $U\circ i_2$ too
has to be of the form described in
Proposition~\ref{Prop:restrictionpsl2r},
hence $U((i_1 (H)-i_2 (H))i(T))$ is a scalar. The rest of the statement is now
obvious. \qed
\begin{Cor}
Let $I\to\cA(I)$ be a second quantization conformal precosheaf on
$S^1$ as
described in the following subsection, $I\to\cA^d(I)$ be its dual net
and $U$ be the above representation of
$\PSL*_\Pnot\PSL$, then the
irreducible components of $U$ belongs to the family described in
Theorem~\ref{freeprod}.
\end{Cor}
\begin{proof}
Since the dual net may be described in terms of the local algebras
$\cA^d(a,b)$ $\doteq $ $\cA(-\infty,b)\cap\cA(a,\infty)$ and the map which
associates
a local algebra with a local subspace is an isomorphism of complemented lattices
(cf. \cite{[Araki]}), the representation $U$ is indeed a second quantization.
On the one-particle space, the construction of the dual net may be done on any
irreducible component, and the result follows.
\end{proof}
\begin{Cor}\label{Lem:extrepsl2r}
  Every irreducible lowest weight representation of $\PSL$ extends to
  a (anti-)representation of $\Mob$ in a unique
  (up to a phase) way.
\end{Cor}
\begin{proof}
Let $E_\lambda$, $T_\lambda$ and $S_\lambda$ be the generators of the
representation of  lowest weight $\alpha$ as above. Since $\Mob$ is generated
by $\PSL$ and e.g. the matrix
$\left(\begin{matrix}-1&0\cr0&1\cr\end{matrix}\right)$, which correspond to the
change of sign on $\Reali$, we need an antiunitary $C$ which satisfies
$CE_\lambda C=E_\lambda$, $CT_\lambda C=-T_\lambda$ and $C S_\lambda
C=-S_\lambda$. (Because $E_{\lambda}, T_{\lambda}$ and $S_{\lambda}$ are
generators, C is then uniquely defined up to a phase.)
  Since in the Schr\"odinger representation the complex
conjugation $C$ satisfies the mentioned commutation relations with $T_0$, $S_0$
and $E_0$, it trivially has the prescribed commutation relations with
$T_\lambda$ and $E_\lambda$, and the last relation follows by the formula
$S_\lambda=S_0-\lambda T_0^{-1}$.
\end{proof}
\subsection{A modular construction of free conformal fields on
$S^1$}
\label{subsec:freefields}
For a certain class of pairs $(M,G)$, where $M$ ifs a homogeneous
space for the
symmetry group $G$, modular theory
may be used to construct a net of local algebras on $M$ starting from
a suitable (anti-)
representation of the symmetry group $G$ \cite{[BGL3]}.
For related works, pointing also to other directions, the interested
reader should consult \cite{[Schr1], [Schr2], [Schr3]}.
 We sketch here
the case of the action of the M\"obius group on $S^1$.

We recall that a real subspace $\cK$ of a complex Hilbert space $\cH$ is called
{\sl standard} if $\cK\cap i\cK=\{0\}$ and $\cK+i\cK$ is dense in $\cH$, and
Tomita operators $j$, $\delta$ are canonically associated with any standard
space (cf. \cite{[RvD]}). One may easily show that the subspaces $\cK'$,
$i\cK$ and $i\cK'$ are standard subspaces if $\cK$ is such, where the symplectic
complement $\cK'$ is defined by $\cK'=\{h\in\cH\ ;\ \mbox{Im}(h,g)=0\ \forall
g\in\cK\}$.

As shown in \cite{[BGL3]}, with any positive energy
representation of $\Mob$ on a Hilbert space $\cH$ we may uniquely
associate a
family $I\to\cK(I)$ of standard subspaces attached to proper intervals $I$
in $S^1$ satisfying the following properties:
\begin{align*}
1)&\ I_1\subset I_2\Rightarrow \cK(I_1)\subset\cK(I_2)\qquad
&{\text{(isotony)}}\cr
2)&\ \cK(I)'=\cK(I')\qquad&{\text{(duality)}}\cr
3)&\ U(g)\cK(I)=\cK(gI)\qquad&{\text{(conformal covariance)}}\cr
4)&\ \delta_I^{it}=U(\Lambda_I(t)),\ j_I= U(r_I) \qquad
&{\text {(Bisognano-Wichmann property)}}\cr
\end{align*}
that is to say, $I\to\cK(I)$ is a local conformal precosheaf of standard
subspaces of $\cH$ on the proper intervals of $S^1$.
The subspaces $\cK(I)$ are defined as
$$
\cK(I)\doteq \{h\in \cH \, |\, j_I\delta_I^{\frac{1}{2}}h=h\}.
$$
We notice that the
precosheaf is irreducible, i.e. $\vee\cK(I)$ is dense in $\cH$, if and only if
$U$ does not contain the trivial representation.
Applying the second quantization functor, we then get a local conformal
precosheaf of von~Neumann algebras acting on the Fock space $e^{\cH}$.

Now we observe that we may extend the lowest weight representations (with
integral $\alpha=n$) described in Proposition~\ref{Prop:restrictionpsl2r},
to an (anti-)representation of $\Mob$ (cf. Corollary~\ref{Lem:extrepsl2r})
in a coherent way, e.g. choosing the complex conjugation $C$ for any
$\alpha$ as in the proof of Corollary~\ref{Lem:extrepsl2r}, and we get a
family of  conformal precosheaves $I\to\cK_n(I)$ of standard spaces on
$S^1$.  The groups $e^{tE}$ and  $e^{tT}$ have a unique fixed point,
namely $\{\infty\}$, in $S^1$ and we may therefore identify
$S^1\setminus\{\infty\}$ with $\Reali$. Then, since the modular groups
of the half-lines do not depend on $n$ by construction, we get
$\cK_{n}(I)=\cK_{1}(I)$ when $I$ is a half-line, namely $\{\infty\}$ is one of
its edges.
\begin{Thm}
  With the above notations, let $I\to\cK(I)$ be a conformal precosheaf of
standard subspaces of $\cH$
  on $S^1$ such that $\cK(I)=\cK_{1}(I)$ for any half-line $I$. Then,
there exists $n\in\Naturali$ such that $\cK(I)=\cK_{n}(I)$ for any
interval $I$.
\end{Thm}
\begin{proof}
  By the Bisognano-Wichmann Theorem for conformal precosheaves on $S^1$
  (cf. \cite{[BGL],[Ga-Fr]})  we get that
  $U\! |_{\Pnot}$ coincides with
  the restriction to $\Pnot$ of the $n=1$ lowest weight
  representation of $\covPSL$.
  Therefore, since $U$ is a positive energy representation, it should be
  of the form described in Proposition~\ref{Prop:restrictionpsl2r}.
\end{proof}
Suppose now we start with the unique irreducible positive energy unitary
representation $U$ of  the translation-dilation group, with non-trivial
restriction to the  translation subgroup, on a Hilbert space $\cH$.
According to \cite{[BGL3]}, we may then consider the associated precosheaf of
standard subspaces $I\to\cK(I)$ on the half-lines $I\subset\Reali$.
The following Corollary summarizes some properties discussed in
Section~\ref{sec:first} and some results of the previous subsection.
\begin{Cor}\label{Cor:netext}
Let $I\to\cK(I)$ the above described precosheaf  on the half-lines of $\Reali$.
 Then there exists a bijective correspondence between
\begin{itemize}
\item Extensions of $\cK$ to a  local conformal precosheaf on the
intervals of $S^{1}$.
\item Real standard subspaces of $\cK(-\infty,1)'\cap\cK(0,\infty)$
-halfsided invariant w. r. t. the subgroup of dilations centered
in $0$ and +halfsided invariant w.r.t. the subgroup of dilations centered
in $1$.
\item The real linear spaces $\cK_n(0,1)$, $n\in\Naturali$.
\end{itemize}
\end{Cor}
\subsection{Multiplicative perturbations: a formula for the canonical
endomorphism}\label{subsec:canendo}

We now give an alternative way to pass from the representation of  lowest
weight 1 to the representation with lowest weight $\alpha\geq 1$. In this
subsection we denote by $E,T,S$ the Lie algebra generators in the lowest
weight 1 representation, and with $E,T,S_\alpha$ the corresponding
generators in the lowest weight $\alpha$ case.
Instead of defining the generator $S_{\alpha}$ as $S - \lambda  T^{-1}$,
$\lambda =\alpha(\alpha -1)$, we will define  the unitary
$R_{\alpha}$ corresponding to the ray inversion or, equivalently, the unitary
\begin{equation}\label{gammadef}
\gamma=\gamma_{\alpha}=R_{\alpha}R=J_{\alpha}J
\end{equation}
where $J$, resp. $J_{\alpha}$ is the modular conjugation of $\cK(-1,1)$,
resp. $\cK_{\alpha}(-1,1)$, as $J=CR$ and $J_{\alpha}=CR_{\alpha}$ with the
same anti-unitary  conjugation commuting with them. In the examples
below the second
quantization of $\gamma$ will implement the canonical  endomorphism of the
inclusion of algebras $\cA_{\alpha}(-1,1)\subset\cA(-1,1)$ given by
$(\alpha-1)$-derivative of the current algebra (in case $\alpha$  is an
integer).

We now make some formal motivation calculations, that may however be given a
rigorous meaning. First note that $\gamma$ commutes with $E$, because both
$J$ and
$J_{\alpha}$ commute with $E$, hence $\gamma$ must be a bounded Borel
function of
$E$
\begin{equation}
\gamma=f_{\alpha}(E)
\end{equation}
because the bounded Borel functions of $E$ form a maximal abelian
von Neumann algebra.

In order to determine $f=f_{\alpha}$, note that the formulas
\begin{gather}
RTR=S\\
R_{\alpha}TR_{\alpha}=S_{\alpha}=S-\lambda T^{-1}
\end{gather}
implies $\gamma^* T\gamma=T-\lambda RT^{-1}R$, hence
\begin{equation}\label{XXX}
\gamma^* T\gamma T^{-1}=1-\lambda RT^{-1}RT^{-1}=1-\lambda (TRTR )^{-1}.
\end{equation}
On the other hand
\begin{equation}
TRTR=TS= E(E-1)
\end{equation}
by (\ref{Casimir}), and since $TET^{-1}=E-1$, thus
\begin{equation}
Tf(E)T^{-1}=f(E-1),
\end{equation}
formula (\ref{XXX}) implies $f$ to satisfy the functional equation
\begin{equation}
\frac{f(z-1)}{f(z)}=1- \frac{\lambda}{z(z-1)}
\end{equation}
and $|f(z)|=1$ for all $z\in i\Reali$.
\begin{Prop}\label{Prop:canendo}
If $\alpha= n$ is an integer, then
\begin{equation}\label{gamma}
\gamma =\frac{(E-1)(E-2)\cdots (E-n+1)}{(E+1)(E+2)\cdots (E+n-1)}.
\end{equation}
In the general case, $\gamma=f_\alpha(E)$ with
\begin{equation}\label{gamma2}
f_{\alpha}(z)=\frac{\Gamma(z+1)\Gamma(z)}{\Gamma(z+\alpha)\Gamma(z-\alpha+1)}
\end{equation}
where $\Gamma$ is the Euler Gamma-function.
\end{Prop}
\begin{proof} Let $\gamma_{\alpha}$ be given by the formula
(\ref{gamma}). In order to check that $\gamma_{\alpha}$
gives (up to a phase) the unitary (\ref{gammadef}) it is enough to check
that
\begin{gather}
\gamma_{\alpha}E\gamma_{\alpha}^*=R_{\alpha}ER_{\alpha}=E\\
\gamma_{\alpha}S\gamma_{\alpha}^*=R_{\alpha}TR_{\alpha}=S_{\alpha}
\end{gather}
because the representation generated by $E$ and $S$ is irreducible,
see
(\ref{Casimir}) and the remarks below it.

The first equation is obvious because $\gamma_{\alpha}$ is a
function of $E$. To verify the second equation we notice that from
$S=RTR$ we get S positive and non-singular and (\ref{Casimir})
shows that this also holds for $E(E+1)=TS$. Since $SES^{-1}=E+1$ the
functional equation for $f_{\alpha}$ implies
$$
f_{\alpha}(E)Sf_{\alpha}(E)^*=f_{\alpha}(E)f_{\alpha}(E+1)^*S=
(1-\frac{\lambda}{E(E+1)})S=S-\lambda T^{-1}=S_{\alpha}.
$$
\end{proof}
\subsection{Lowest weight representations of $\PSL$ and derivatives of the
$U(1)$-current }\label{sec:currents}
 On the space $\cinfcerchio$ of real valued smooth functions on the circle
$S^1$, we consider the seminorm
\begin{equation*}
  \|\phi\|^2 =\sum_{k=1}^{\infty} k |\fphi_k|^2
\end{equation*}
and the operator $\cI$ : $\widehat{\cI\phi}_k=-i \mbox{sign}(k)\fphi_k$, where
the $\hat\phi_k$'s denote the Fourier coefficients of $\phi$.

Since $\cI^2=-1$ and $\cI$ is an isometry w.r.t. $\|\cdot\|$,
$(\cinfcerchio,\cI,\|\cdot\|)$ becomes a complex vector space with a
positive bilinear form, defined by polarization. Thus, taking the quotient
by constant functions and completing, we get a complex Hilbert space
$\cH$.

We note that the symplectic form $\omega$ may be written as
\begin{equation*}
  \omega(f,g)=\mbox{Im}(f,g)=\frac{-i}{2}\sum_{k\in\Interi}
  k\hat{f}_{-k}\hat{g}_k=\frac{1}{2}\int_{S^1} g df .
\end{equation*}
One might recognize this form as coming from the commutation relations
for $U(1)-$currents.
 The natural action of $\PSL$ on $S^1$ gives rise to a unitary representation
on $\cH$:
\begin{equation*}
  U(g)\phi(t)=\phi(g^{-1}t)
\end{equation*}
Then, observing that $\cI\cos kt=\sin kt$ for $k \ge 1,$ it is easy to see
that $\cos kt$ is
an eigenvector of the rotation subgroup $U(\theta)$:
\begin{equation*}
  U(\theta)\cos kt=\cos k(t-\theta)=(\cos k\theta + \sin k\theta\cI) \cos kt =
  e^{i k\theta}\cos kt, \qquad k\ge 1
\end{equation*}
and that all the eigenvectors have this form. Therefore the representation has
lowest weight $1$.

We need another description of the Hilbert space $\cH$ which is more suitable
to be generalized. First we choose another coordinate on $S^1$, namely
$x=\tan(t/2)$, $x\in\Reali$, and therefore identify $\cinfcerchio$
with $\cC^\infty(\dot{\Reali})$, $\dot{\Reali}$ being the one-point
compactification of $\Reali$. Since the symplectic form is the integral of a
differential form it does not depend on the coordinate:
 $$
\omega(f,g)=\frac{1}{2}\int_\Reali g(x)df(x)
 $$
 A computation shows that the anti-unitary $\cI$ applied to a function $f$
coincides up to an additive constant with the convolution of $f$ with the
distribution
$1/(x+i0)$ on $\Reali$, therefore, since the symplectic form is trivial on the
constants, the (real) scalar product may be written as
\begin{equation}\label{one-norm}
\begin{split}
\langle f,g\rangle
&=\omega(f,\cI g)=\frac12\int\left(\frac1{x+i0}*g(x)\right)f'(x)dx\\
&=\frac1{4\pi}\int f(x)g(y)\frac1{|x-y+i0|^2}dxdy
=const\int_0^\infty p\hat f(-p)\hat g(p)dp
\end{split}
\end{equation}
and $\cH$ may be identified with the completion of $\cC^\infty(\dot\Reali)$
w.r.t. this norm.

Note that since $\cI f =-if$ if supp$\hat{f}\subset [0,+\infty)$,
$\cH$ is also the completion of $C^{\infty}(\mathbb R,\mathbb C)$
modulo $\{f | \hat{f}|_{(-\infty,0]}=0\}$ with scalar product
$(f,g)=\int_0^{\infty}p\overline{\hat f(p)}\hat{g}(p)dp$.

Let us now consider the space  $X^n\doteq
\cC^\infty(\dot\Reali)+\Reali^{2(n-1)}[x]$,
$n\geq1$, where $\Reali^p[x]$ denotes the space of real polynomials of degree
$p$, and the bilinear form on it given by
 $$
\langle f,g\rangle_n=\frac1{4\pi}\int f(x)g(y)\frac1{|x-y+i0|^{2n}}dxdy
 $$
 It turns out that $\langle\ \cdot\ ,\ \cdot\ \rangle_n$ is a well defined
positive semi-definite bilinear form on $X^n$ which degenerates exactly on
$\Reali^{2(n-1)}[x]$. On this space one may define also a symplectic form by
 $$
\omega_n(f,g)=\frac12\int f(x)g(y)\delta_0^{(2n-1)}(x-y)dxdy
 $$
This form might be read as the restriction of $\omega_1$ to the
n-th derivatives. Therefore we can recognize this symplectic form as
coming from the commutation relations for the n-th derivatives of
$U(1)-$currents. This form again
degenerates exactly on $\Reali^{2(n-1)}[x]$, and the operator $\cI$
defined before connects the positive form with the symplectic form for any $n$
in such a way that
$(\cdot,\cdot)_n\doteq \langle\cdot,\cdot\rangle_n+i\omega_n(\cdot,\cdot)$
becomes a
complex bilinear form on $(X^n,\cI)$.
We shall denote by $\cH^n$ the complex Hilbert space obtained by completing the
quotient $X^n/\Reali^{2(n-1)}[x]$.

With any matrix $g=\left(\begin{matrix}a&b\cr c&d\cr\end{matrix}\right)$ in
${\mbox{\tt SL}(2,\Reali)}$ we may associate the rational transformation
$x\to gx=\frac{ax+b}{cx+d}$ and then, for any $n\geq1$, the operators
$U^n(g)$ on $X^n$:

 $$
U^n(g)f(x)=(cx-a)^{2(n-1)}f(g^{-1}x).
 $$
It turns out that $g\to U^n(g)$ is a representation of $\PSL$, $n\geq1$, and
that the positive form is preserved (cf. \cite{[Y]}) as well as the symplectic
form and the operator $\cI$, therefore $U^n$ extends to a unitary
representation of $\PSL$ on $\cH^n$.

We remark that while $X^n$ and $\Reali^{2(n-1)}[x]$ are globally preserved by
$U^n$, the space $\cC^\infty(\dot\Reali)$ is not, and that explains why the
space $X^n$ had to be introduced.

By definition the space $\cH^1$ coincides with the space $\cH$
and the representation $U^1$ with the representation $U$, which we
proved to be lowest weight 1.
We observe that, for functions in $\cC^\infty(\dot\Reali)$, one gets
\begin{align*}
\langle f,g\rangle_n&=\frac12\int_\Reali|p|^{2n-1}\hat f(-p)\hat g(p)\cr
\omega(f,g)_n&=\frac12\int_\Reali p^{2n-1}\hat f(-p)\hat g(p)\cr
\end{align*}
hence
 $$
(f,g)_n=(D^{n-1}f,D^{n-1}g)_1,
 $$
i.e. $D^{n-1}$ is a unitary between $\cH^n$ and $\cH^1\equiv\cH,$
where D is the derivative operator.
The following holds:
\begin{Thm}
The representation $U^n$ has lowest weight $n$.
\end{Thm}
\begin{proof}
Making use of the results of Proposition~\ref{Prop:canendo}, we have to show
that
 $$
R_nR=\prod_{k=1}^{n-1}\left(\frac{E-k}{E+k}\right),\qquad n\geq1
 $$
where  $R_n=D^{n-1}U^n(r)(D^{n-1})^*$
  with $r$ the ray
inversion, $R=R_1.$  This amounts to prove
\begin{equation}\label{indhp}
D^{n-1}U^n(r)=
\prod_{k=1}^{n-1}\left(\frac{E-k}{E+k}\right) U(r)D^{n-1}.
\end{equation}
Now we take equation~(\ref{indhp}) as an inductive hypothesis. Then,
equation(~\ref{indhp}) for $n+1$ can be rewritten, using
the inductive hypothesis and the relation $U^{n+1}(r)=x^2U^n(r)$, as
\begin{equation}\label{indth}
D^n (x^2 U^n(r))=\left(\frac{E-n}{E+n}\right) D^{n-1}U^n(r)D.
\end{equation}
Finally we observe that $U^n(r)D=x^2DU^n-2(n-1)xU^n$, hence
equation~(\ref{indth}) is equivalent to
\begin{equation}\label{indth2}
(E+n)D^n(x^2\cdot)=(E-n)D^{n-1}(x^2D\cdot-2(n-1)x\cdot).
\end{equation}
Since $E=-xD$, equation~(\ref{indth2}) follows by a straightforward
computation.
\end{proof}

\begin{Prop}\label{dunitary}
The unitary representations of $\PSL$ on $\cH$ given by
 $$
D^{n-1}U^n (D^{n-1})^*,\qquad n\geq1
 $$
coincide when restricted to the subgroup of translations and dilations on
$\Reali$.
\end{Prop}
\begin{proof}
We have to prove that $D^{n-1}U^n(g)=U(g)D^{n-1}$ when $g$ is a translation or a
dilation. For translations, $U^n(t)f(x)=f(x-t)$, and the equality is obvious;
for dilations,
$g=\left(\begin{matrix}
e^{\lambda/2}&0\cr 0&e^{-\lambda/2}\cr
\end{matrix}\right)$, $U^n(\lambda)f(x)=e^{\lambda n}f(e^{-\lambda}x)$, hence
$D^{n-1}U^n(\lambda)f(x)=f^{(n)}(e^{-\lambda}x)=U(\lambda)D^nf(x)$.
\end{proof}
The family of representations $D^{n-1}U^n (D^{n-1})^*$ on the Hilbert
space $\cH$ constitute a concrete realization of the family of (integral)
lowest weight representations
described in Proposition~\ref{Prop:restrictionpsl2r}, therefore we may
construct a family of local conformal precosheaves of standard subspaces of
$\cH$ as explained in Subsection~\ref{subsec:freefields}. In the next
subsection we shall give another description of these precosheaves,
showing that they coincide with the ones described in \cite{[Y]}.
\subsection{Relations among local spaces}\label{regular}
Let us fix an $n\geq1$ and, for any proper interval $I$ of $\dot\Reali$,
let us set
 $$
X^n(I)=\{f\in X^n:f|_{I'}\equiv0\}.
 $$
It is easy to check that these spaces satisfy the properties
\begin{align*}
  1.\ &I_1\subset I_2\Longrightarrow X^n(I_1)\subset X^n(I_2)\quad
    \mbox{(isotony)},\\
  2.\ &I_1\cap I_2=\emptyset\Longrightarrow X^n(I_1)\subset X^n(I_2)'
    \quad \mbox{(locality)},\\
  3.\ &U^n(g)X^n(I)=X^n(gI), \forall g\in\PSL\quad \mbox{(covariance)},
\end{align*}
and that the immersion $i_n^I:X^n(I)\to\cH^n$ is injective. Therefore the
spaces $\cK^n(I)\doteq (i_n^IX^n(I))^-,$ where the closure is taken
w.r.t. $\| \cdot \|_n,$ form a local conformal
precosheaf of subspaces of $\cH^n$, and the following property obviously holds:
\begin{align*}
4.\ &\bigvee_{I\subset \dot\Reali}\cK^n(I)=\cH^n\quad\mbox{(irreducibility)}.
\end{align*}
Therefore, by the first quantization version  of results mentioned in
Section~\ref{sec:first}, these spaces are standard, the Bisognano-Wichmann
property
and duality on the circle hold.

Now we identify $\cH^n$ with $\cH$ via the unitary $D^{n-1}$, and set
$\cK_n(I)\doteq D^{n-1}\cK^n(I)$. Then, if $I$ is compact in $\Reali$ and
$f\in\cK_n(I)$, $f$ may be integrated $n-1$ times, giving a function which has
still support in $I$, therefore
\begin{equation}
\cK_n(I)=\{[f]\in\cH:f|_{I'}=0,\ \int t^{j}f =0,\
j=0,\dots,n-2\},
\quad I\subset\subset\Reali\tag{a}
\end{equation}
where $[f]$ denotes the equivalence class of $f$ modulo polynomials.

If $I$ is a half line in $\Reali$, $\cK_n(I)$ is
an invariant subspace of the dilation subgroup, which is the modular group of
$\cK(I)$.
Using Takesaki's result, see \cite{[Str-Zs]}, this implies that
\begin{equation}
\cK_n(I)=\cK(I),
\qquad
I\, \mbox {a half-line},\tag{b}
\end{equation}
(For an alternative proof of this fact, see \cite{[Y]}.)

Then, by duality and the formula for the compact case, we obtain
\begin{equation}
\cK_n(I)=\{[f]\in\cH:f|_{I'}=p_{f,I}\in\Reali^{n-1}[x]\}
\quad
I'\subset\subset\Reali .\tag{c}
\end{equation}
Finally we observe that, since Bisognano-Wichmann property holds, these
precosheaves coincide with those abstractly constructed in
Subsection~\ref{subsec:freefields}.

Now, we fix a bounded interval in $\Reali$,
e.g. $(-1,1)$, and consider the family $\cK_{n}\doteq \cK_{n}((-1,1))$. The
concrete characterization of $\cK_n$ given in the preceding subsection shows
that $\cK_m\subseteq\cK_n$ if $m\ge n$. Now, we may show
\begin{Thm}\label{Thm:dimecodim}
  The following dimensional relations hold:
  \begin{align*}
    &\mbox{codim}(\cK_{m}\subset\cK_{n})=m-n,\qquad m\ge n,\\
    &\mbox{dim}(\cK'_{m}\cap\cK_{n})=\max((m-n-1), 0).
  \end{align*}
\end{Thm}
Before proving Theorem~\ref{Thm:dimecodim}, we discuss some of its
consequences.
\begin{Dfn} A precosheaf $\cK$ is said $n-${\it regular} if, for any
partition of $S^1$ into
  $n$ intervals $I_1,\dots,I_n$, the linear space
$\bigvee_{j=1}^{n}\cK(I_j)$ is dense in $\cH$.
\end{Dfn}
We recall that irreducible conformal precosheaves are 2-regular, because
duality  holds and local algebras are factors.
\begin{Cor}\label{Cor:regularity} The conformal precosheaf $\cK_1$ is
$n$-regular for any $n$.
  The conformal precosheaf $\cK_2$ is $3$-regular but it is not $4$-regular.
  The conformal precosheaves $\cK_n$, $n\ge 3$, are not $3$-regular.
  Moreover, strong additivity and duality on the line hold for the precosheaf
  $I\to\cK_1(I)$ only, therefore it is the dual precosheaf of $I\to\cK_n(I)$
  for any $n$.
\end{Cor}
\begin{proof}
First we recall that a precosheaf is strongly additive if and only if it
coincides with its dual precosheaf. Then, the precosheaf $\cK\equiv\cK_1$ is
strongly additive because its dual net should be of the form $\cK_n$  (cf.
Corollary~\ref{Cor:netext}) and should satisfy $\cK^d(-1,1)\supseteq\cK(-1,1)$.
As a consequence, $\cK$ is  $n$-regular for any  $n$.

Then, since the spaces for the half-lines do not depend on $n$, the dual
net of
$\cK_n$ does not depend on $n$ either, hence coincides with $\cK$.

Since $\PSL$ acts transitively on the triples of distinct points, we may study
$3$-regularity for the special triple $(-1,1,\infty)$ in
$\Reali\cup\{\infty\}$. Then,
  \begin{equation*}
    \begin{split}
     &(\cK_{n}(\infty,-1)\vee\cK_{n}(-1,1)\vee\cK_{n}(1,\infty))'\\
     &\qquad\qquad =
       (\cK_{1}(\infty,-1)\vee\cK_{n}(-1,1)\vee\cK_{1}(1,\infty))'\\
     &\qquad\qquad = (\cK_{1}(-1,1)'\vee\cK_{n}(-1,1))'
     = \cK_{n}(-1,1)'\wedge\cK_{1}(-1,1)
    \end{split}
\end{equation*}
  where we used strong additivity and duality for $\cK_1$.
  By Theorem~\ref{Thm:dimecodim}, $3$-regularity holds if and only if $n=1,2$.

  Violation of $4$-regularity for $\cK_2$ may be proved by exhibiting a
  function which is localized in the complement of any of the intervals
$(\infty,-1)$, $(-1,0)$, $(0,1)$, $(1,\infty)$, i.e. belongs to
$\cK_2(-1,0)'\cap\cK_{2}(0,1)'\cap \cK_{1}(-1,1)$:
  \begin{equation*}
    \phi(x)=\left\{\begin{array}{ll}
                    1+x &\mbox{if}\quad -1\geq x \geq 0,\\
                    1-x &\mbox{if}\quad  0\geq x \geq 1,\\
                     0  &\mbox{if}\quad       |x|\geq 1
                   \end{array}
            \right.
  \end{equation*}
  In the same way we may construct a function which violates $3$-regularity for
  $\cK_3$, namely
  \begin{equation*}
    \phi(x)=\left\{\begin{array}{ll}
                    x^2 -1 &\mbox{if}\quad |x|<1,\\
                     0     &\mbox{if}\quad |x|\ge 1
                   \end{array}
            \right.
  \end{equation*}
  Clearly, $\phi\in \cK_3(\infty,-1)'\cap\cK_3(-1,1)'\cap \cK_3(1,\infty)'=
\cK_{3}'\cap\cK_{1}$.
\end{proof}
\begin{Lemma}\label{lemma:codimensioni}
  $\mbox{codim}(\cK_{m+1}\subset\cK_{m})=1$.
\end{Lemma}
\begin{proof}
  Since $\cK_{m+1}=\{\phi\in\cK_{m}\ :\ \int x^{m-1}\phi(x)dx =0\}$, and we
  may find a function $\psi_{m-2}\in\cinfcom (\Reali)\ $ :
$\psi^{'}_{m-1}(x)=x^{m-1}$,
  $x\in (-1,1)$, we get $\cK_{m+1}=\{\phi\in\cK_{m}\ :\
  \omega(\psi_{m-1},\phi)=0\}$. Because the functional
$\phi\longrightarrow\omega
  (\psi_{m},\phi)$ is continuous and non zero on $\cK_{m}$, the thesis follows.
\end{proof}
{\bf Proof of Theorem~\ref{Thm:dimecodim}:}
The first statement of the Theorem easily  follows from
Lemma~\ref{lemma:codimensioni}. Now, let us consider the  relative commutants
$\cK_{m+p}'\cap\cK_{m}$. We observe that, by Poincar\'e inequality, the norm on
$\cK_1$ is equivalent to the Sobolev norm for the space $H^{1/2}$, i.e., we
may identify $\cK_{1} \simeq H^{1/2}(-1,1)$ as real Hilbert spaces. We also
recall that the Dirac measure $\delta$ does not belong to $H^{-1/2}$, but
belongs to $H^{-1/2-\epsilon}$ for each $\epsilon>0$ (see, e.g.,
\cite{[Treves]}). Then
  \begin{align*}
    \cK_{m+p}'\cap\cK_{m}&=\{\phi\in\cK_{m}\ :\ \langle \phi',\psi\rangle =0 \
    \forall\psi\in\cK_{m+p}\}\\
    &=\{\phi\in H^{1/2}(-1,1)\ :\  \int t^{j}\phi =0,\ j=0,\dots,m-2,\\
    &\qquad\qquad\qquad \langle \phi^{(m+p)},\psi\rangle =0,\
                         \forall\psi\in H^{m+p-1/2}(-1,1)\}.
  \end{align*}
Then $f\doteq \phi^{(m+p)}\in H^{1/2-m-p}\{-1,1\}$, i.e., $f$ should
  be a combination
  of Dirac's $\delta$ measures with supports in $\{-1,1\}$ and their
  derivatives.
  Since $f\in H^{1/2-m-p}$, it has the form
  $f=\sum_{j=0}^{m+p-2} (c_{j} \delta^{(j)}_{(-1)} + d_{j}\delta^{(j)}_{(1)})$.
  The condition $\phi\in\cK_{m}$ may be written as
  \begin{equation}\label{eq:condition}
    \langle f,t^{q}\rangle =0,\qquad q=0,\dots, 2m+p-2.
  \end{equation}
  The dimension of $\cK_{m+p}'\cap\cK_{m}$ will be the difference between
  the dimension of the space
  $\Delta\doteq \{\sum_{j=0}^{m+p-2}(c_{j} \delta^{(j)}_{(-1)} +
  d_{j}\delta^{(j)}_{(1)})\ :\ c_{j},d_{j}\in\Reali\}$, which is $2(m+p-1)$, and
  the number of independent conditions in equation (\ref{eq:condition}).
  We may also write the conditions in equation (\ref{eq:condition}) as
  \begin{equation*}
    \langle f,P\rangle =0\qquad \mbox{where\ }\  P\ \mbox{is a polynomial
    of degree\ } 2m+p-2.
  \end{equation*}

They are independent if the only polynomial $P$ of degree $\le 2m+p-2$
satisfying $\langle f,P\rangle =0$ for any $f\in\Delta$ is the null polynomial.
Indeed, such polynomial should have zeroes with multiplicities greater than
$m+p-1$ for the points $-1$ and $1$, therefore, either $p=0$, and then there
exists  exactly one non trivial such polynomial, or $p>0$, and the null
polynomial is the unique solution. In conclusion, if $p>0$, the conditions in
equation (\ref{eq:condition}) are independent, and the dimension of
$\cK_{m+p}'\cap \cK_{m}$ is
  \begin{equation*}
    2(m+p-1)-(2m+p-1)=p-1.
  \end{equation*}
If $p=0$ the independent conditions in equation (\ref{eq:condition}) are
$2m-2$,  and the dimension is $2(m-1)-(2m-2)=0$, which corresponds to the
general
fact  (see, \cite{[GuLo2]}) that local algebras of irreducible conformal
theories
are factors.\qed\medskip

\section{Examples of superselection sectors for the first derivative of the
$U(1)$-current}\label{sec:fourth}

In this section we shall discuss examples of superselection sectors
of the first-derivative  theory. All these sectors are abelian, i.e. are
equivalence classes of automorphisms, and we will see that they are non
covariant under the conformal group. In particular, recalling that the
first-derivative precosheaf is 3-regular but not 4-regular, this shows that the
assumption of 4-regularity in \cite{[Gu-Lo]} cannot be avoided in general in
order to obtain the automatic covariance of superselection sectors.

As we shall see, all these sectors will be obtained by generalizing
methods of the Buchholz-Mack-Todorov approach to sectors (see
\cite{[BMT]}).

On the one hand the conformal net on $\Reali$ associated with the current
algebra contains as a subnet the one associated with the first
($n$-th) derivative of the current algebra (cf.
Corollary~\ref{Cor:regularity}
and the Remark after Corollary~\ref{confdual}), therefore BMT sectors
may be
restricted to the conformal net on
$\Reali$ associated with the first ($n$-th) derivative of the current
algebra. The sectors described here will be extensions (of such
restrictions) to the conformal precosheaf on $S^1$ associated with the
first ($n$-th) derivative of the current algebra.

On the other hand they may be seen as sectors on a suitable global algebra
in a way which is formally identical to the BMT procedure.

Now let $\cA$ be a local conformal precosheaf on $\Reali$ and $\cA^{d}$
its Bisognano-Wichmann dual net.  Consider the unitary $\Gamma=JJ_{d}$,
where $J$ and $J_{d}$ are the modular conjugations of $\cA(-1,1)$ and
$\cA^{d}(-1,1)$ with respect to the vacuum vector $\Omega$, in other
words $\Gamma$ is the product of the two ray inversion unitaries of the
nets $\cA$ and $\cA^{d}$. The unitary $\Gamma$ implements the {\it
canonical endomorphism} $\gamma$ of $\cA^{d}(-1,1)$ into $\cA(-1,1)$.

Let now $\rho$ be a morphism of $\cal A$. We define the ``extension'' of $\rho$
to ${\cal A}^d$ by
$$\tilde\rho=\mbox{Ad}\Gamma^{*}\rho\gamma\ .$$

If $I$ contains the origin (possibly at the boundary), $\gamma$ sends ${\cal
A}^d(I)$ into ${\cal A}(I)$:

\begin{equation}
\mbox{Ad}\Gamma(\cA^{d}(I))=J\cA^{d}(\hat I)'J\subset
J\cA(\hat I)'J=\cA(I)
\end{equation}
where $\hat I$ is the image of $I$ under the ray inversion map.

Therefore, if $\{\rho^I\}$ is the family of
representations defining $\rho$ and $I$ is an interval containing the origin,
then $\tilde{\rho}^I=\gamma^{-1}\rho^I\gamma$ gives a representation of ${\cal
A}(I)$ and these representations are coherent.  However, if $I$ does not
contain the origin, there are two minimal intervals containing both $I$ and the
origin, one in which they are in clockwise order and the other in which they are
in the counterclockwise one, and the two corresponding  representations not
necessarily agree on the algebra of the intersection.

If they do not, $\tilde{\rho}$ is not a representation of the dual precosheaf,
nevertheless, if the point at infinity is removed and $\rho$ is localized in a
compact interval, only one choice remains, and we get a representation of the
net ${\cal A}_0^d$ on the line. Clearly equivalent endomorphisms give rise to
equivalent representations and if we choose the localization region $I_0$ not
containing the origin, say $I_0=(a,b)$, $b>a>0$, then $\tilde{\rho}$ is
localized in $(a,\infty)$.  We have therfore shown that any transportable
sector on ${\cal A}(I)$ gives rise to a (possibly solitonic) sector on ${\cal
A}_0^d(I)$.  (In this lower dimensional theory one might also interpret these
sectors as coming from order variables, \cite{[Schr4]}, Chapter 3.8.)
In subsection \ref{subsec:solitons} we show examples of this phenomenon.

Conversely, if we assume that the two above mentioned representations agree,
we get a representation of the precosheaf ${\cal A}^d$, and assuming again that
the localization interval $I_0$ do not contain the origin, $\tilde{\rho}$ is
localized in $I_0$. If we further assume that $\rho$ is covariant and finite
statistics, we obtain that $\tilde{\rho}$ is finite statistics too,
because the index may be computed by looking at the endomorphisms of
the von Neumann algebra $\cA^d(0,\infty)=\cA(0,\infty)$. Hence $\tilde{\rho}$
is covariant by the strong additivity of the dual net (see \cite{[Gu-Lo]})
and this implies that $\rho$ and $\tilde{\rho}$ determine equivalent
representations of the net ${\cal A}_0$ on the line. In fact, by the
construction of the dual net, the product $J J_d$
of the modular conjugations for the interval $(-1,1)$ relative to the two
theories coincides with the product of the two unitaries implementing the
conformal transformation $t\to -1/t$. Then, denoting by $r$, $r_d$ the
corresponding automorphisms and by $u_d$ and $u$ the unitaries such
that $\mbox{Ad}(u_d) \tilde{\rho} = r_d \tilde{\rho} r_d$ and
$\mbox{Ad}(u)\rho = r\rho r$ we have
$$
\mbox{Ad}(u_d) \tilde{\rho} = r_d \tilde{\rho} r_d = r \rho r =
\mbox{Ad}(u) \rho .
$$

\subsection{Buchholz, Mack and Todorov approach to
sectors of the current algebra }
\label{subsec:BMT}

We defined the one-particle space for the current algebra as the completion of
the space $X=X^1=\cC^\infty(\dot\Reali)$ modulo constant functions w.r.t.
the
norm given in (\ref{one-norm}).
We may  then define $\cA(S^1)$ as the $^*$ - algebra generated by
$W(h), h\in X$ with the relations $W(h)W(h)^*=1$ (unitarity) and
$W(h)W(k)=exp(i/2\omega(h,k))W(h+k)$ (CCR).

BMT automorphisms of $\cA(S^1)$ are then given in terms of differential
forms $\phi$ on $S^1$. Setting
 $$
\alpha_\phi(W(h))\doteq e^{i\int h\phi}W(h)
 $$
it is easy to see that $\alpha$ extends to an automorphism of $\cA(S^1)$.
By CCR, it follows that $\alpha_\phi$ is inner if and only if the form $\phi$
is exact, i.e. there exists a function $f\in X$ s.t. $\phi=df$, and that
two
automorphisms $\alpha_\phi$, $\alpha_\psi$ are equivalent if and only if
$\int\phi=\int\psi$, i.e. if the two forms give the same cohomology class in
$H^1(S^1)$. The constant $Q(\alpha_\phi):=\int\phi$ will be called the charge of
$\alpha_\phi$.

For any open interval $I$ in $S^1$ we set $\cA(I)$ to be the subalgebra of
$\cA(S^1)$ generated by Weyl unitaries $W(h)$ such that the support of $h$ is
contained in $I$. Clearly the algebras associated with disjoint intervals
commute and $\beta_g\cA(I)\doteq \cA(gI)$ where $\beta_g(W(f))\doteq W(U(g)f)$.

We observe that BMT automorphisms are {\sl locally internal}, i.e. for any
interval $I$ and any form $\phi$ there exists a function $f$ with support in
some larger interval $\hat I$ such that $df|_I\equiv\phi|_I$, therefore
$\alpha_\phi|_{\cA(I)}\equiv {\rm{ad}}W(f)|_{\cA(I)}$.

Also, the superselection sectors corresponding to a given charge are
conformally covariant w.r.t. the adjoint action of the conformal group on $X$,
i.e. the automorphisms $\alpha_\phi$ and
$\beta_g\cdot\alpha_\phi\cdot\beta_{g^{-1}}$ are in the same class for any
conformal transformation $g$. Indeed, since the class of inner automorphisms
is globally stable under the action of the conformal group and the charge is
additive, namely  $Q(\alpha_\phi\circ\alpha_\psi)$ $=$ $Q(\alpha_\phi)$ $+$
$Q(\alpha_\psi)$, the action of $\PSL$ on BMT automorphisms gives a
linear action on BMT charges, i.e. a one dimensional linear representation of
$\PSL$. Any such representation being trivial, BMT sectors are covariant.

Now we give a local description for these sectors.
We observe that the second quantization algebra associated with the standard
space $\cK(I)$ coincides with $\pi(\cA(I))''=\cR(I)$, where $\pi$ is
the vacuum representation of $\cA(S^1)$ on the Fock space $e^H$. Moreover,
the map $\pi|_{\cA(I)}$ is faithful, and the restriction of $\alpha_\phi$ to
$\cA(I)$, being implemented in $\cA(\hat I)$, uniquely extends to a normal
automorphism of $\cR(I)$. As a consequence, $\alpha_\phi$
gives rise to a representation  $I\to\alpha^I_\phi$ of the precosheaf $\cA$ in
the sense of \cite{[GuLo2]}, where
 $$
 \alpha^I_\phi(\pi(W(h)))=e^{i\int h_I\phi}W(h)
 $$
and, recalling that $h\in H$ is localized in $I$ if it is equal to a constant
$c_I$ in $I'$, we have set $h_I\doteq h-c_I$.

We described BMT locally normal representations via automorphisms of
$\cA(S^1)$. Conversely, the global algebra $\cA(S^1)$ plays the role of
the universal algebra w.r.t. the family of the locally normal BMT
representations, in the sense that the classes of such representations modulo
unitary equivalence appear as classes of global  automorphisms of $\cA(S^1)$ up
to inners.

\subsection{Restriction of localized sectors}\label{subsec:BMTonOneDerivative}

As we have already seen, local algebras associated with compact intervals on
the line for
the first-derivative net may be described as
 $$
\cR_2(I)\doteq \{W(h)\in\cR(I):\int h(x)dx=0\}''.
 $$
Of course these algebras form a net of local algebras on the real line which is
covariant with respect to the action of translations and dilations, but Haag
duality does not hold on $\Reali$.
The quasi-local algebra $\cA_2(\Reali)$ generated by the algebras of
compact intervals is a subalgebra of the quasi-local algebra $\cA(\Reali)$ of
the current algebra on the line, therefore any BMT automorphism of
$\cA$
 localized in
some compact interval $I$ gives a representation of  $\cA_2(\Reali)$ which
is equivalent to the vacuum representation if and only if it has zero charge,
but, due to the failure of Haag duality, the intertwining unitary is not
necessarily localized in $I$.
Such a unitary exhibits instead a solitonic localization, i.e. it necessarily
belongs to the von~Neumann algebra of any half line containing the localization
region.
The restrictions of BMT sectors are then translation and dilation covariant.

On the contrary, if we consider classes of automorphisms of
$\cA_2(\Reali)$ modulo inners, a new charge appears, i.e. two automorphisms
$\alpha_\phi$ and $\alpha_\psi$ are equivalent if and only if both
$\int\phi=\int\psi$ and $\int t\phi=\int t\psi$ are equal. As a consequence,
such sectors are no longer translation covariant.

\subsection{Conformal solitonic sectors}\label{subsec:solitons}

In the first-derivative theory, the
automorphism $\alpha_\phi$ is localized in a compact interval $I$ of $\Reali$
when $\phi$ is constant outside $I$, therefore solitonic sectors on
$\cA(\Reali)$ may become localized when restricted to $\cA_2(\Reali)$.
This shows that, conversely, sectors on $\cA_2$ may become solitonic when
extended to $\cA$ as described at the beginning of this section.

Here we shall consider $\phi$ as a function on $\Reali$ rather than as a
differential form, identifying $\phi(t)$ with $\phi(t)dt$.
If $\phi$ is constant outside $I$, $\alpha_\phi$ is equivalent to the
vacuum (as a representation) when both $\phi(+\infty)=\phi(-\infty)$
$(=0)$ and $\int_{-\infty}^{+\infty}\phi=0$. As a consequence,
superselection sectors are described by two charges:
 $$
Q_0=\phi(+\infty)-\phi(-\infty)=\int\phi'(t)dt\qquad
Q_1=\int t\phi'(t)dt.
 $$
These sectors are clearly transportable, but a simple computation shows that
they are covariant under translations and dilations if and only if $Q_0=0$,
i.e. only if they are restrictions of BMT sectors.
 Restrictions to $\cA_2(\Reali)$ of solitonic sectors on $\cA(\Reali)$
give then an example of localized non covariant sectors on the line. As we
shall see, these sectors may be extended to transportable sectors on the
circle.

\subsection{Generalized BMT approch to sectors: Local description}
\label{subsec:locsect}
We recall that the conformal precosheaf $\cR_2$ of the first-derivative theory
may be described as second quantization algebras on the same Fock space as the
current algebra, $\cR_2(I)=\{W(h):h\in \cK_2(I)\}''$.

We have seen that
\begin{align*}
\cK_2(I)&=\{[f]\in \cH:f|_{I'}\equiv p_{f,I},
\ p_{f,I}\in\Reali^0[x], \int (f(x)-p_{f,I})dx=0\}
\quad I\subset\subset\Reali\cr
\cK_2(I) &= \{[f]\in \cH: f|_{I'}\equiv p_{f,I}
\ p_{f,I}\in\Reali^0[x]\}
\quad I {\text{ half line}}\cr
\cK_2(I) &= \{[f]\in \cH: f|_{I'}\equiv p_{f,I},
\ p_{f,I}\in\Reali^1[x]\}
\quad I'\subset\subset\Reali.\cr
\end{align*}

In order to extend a BMT automorphism $\alpha_\phi$ to the first-derivative
theory on the circle we have to choose a real number $\lambda$ and then set
 $$
\alpha^I_{\phi,\lambda}(W(f))=
e^{i\langle\phi,f\rangle_{I,\lambda}}W(f)
 $$
where f belongs to $ \cK_2(I),$ with
 $$
\langle\phi,f\rangle_{I,\lambda}=\int\phi(x)(f(x)-p_{f,I}(\lambda))dx.
 $$
Taking $\phi$, $\psi$ such that $Q=\int\phi=\int\psi$ we may compute
$\alpha_{\phi,\lambda}\cdot\alpha^{-1}_{\psi,\mu}$:
\begin{equation}\label{difference}
\begin{split}
(\alpha_{\phi,\lambda}\cdot\alpha^{-1}_{\psi,\mu})^I(W(f))
&=e^{i(\langle\phi,f\rangle_{I,\lambda}-\langle\psi,f\rangle_{I,\mu})}W(f)\\
&=e^{-iQ(\lambda-\mu)\langle\delta'_x,f\rangle}{\text{ad }}
W(h)(W(f)),
\end{split}
\end{equation}
with $ h' = \phi - \psi,$
where $x$ is any point in $I'$.
 Therefore we have proved that two BMT automorphisms with the same (non zero)
charge extend to equivalent automorphisms if and only if $\lambda=\mu$.  Since
a simple calculation shows that the translated automorphism
$\beta_{T(t)}\alpha_{\phi,\lambda}\beta_{T(-t)}$ is equal to
$\alpha_{\phi(\,\cdot\,+t),\lambda-t}$, we conclude that non trivial BMT
sectors give rise to a one parameter family of non covariant sectors on the
circle.

Moreover, when $\lambda\not=\mu$, the automorphism
$\alpha_{\phi,\lambda}\cdot\alpha^{-1}_{\psi,\mu}$ is equivalent to a new
automorphism $\theta_c$, $c=-Q(\lambda-\mu)$:
\begin{equation}\label{theta}
\theta_c^I(W(f))
=e^{ic\langle\delta'_x,f\rangle}W(f),\quad x\in I'.
\end{equation}
 Since $p_{f,I}$ is constant whenever $I\subset\Reali$ and hence
$\langle\delta'_x,f\rangle$ vanishes, we conclude that
$\theta_c$ is localized in only one point, the point at infinity. We may easily
show that $\theta_c$ is invariant under translations and that dilations act on
the charge $c$.

By conjugating $\theta_c$ with a conformal transformation we will get a
family of automorphisms localized in different points of the real line. In
particular, requiring the automorphisms to be localized in zero,
we get a family $\zeta_c$:
\begin{equation}\label{zeta}
\zeta_c(W(f))\doteq e^{ic\int_0^x(f(y)-p_{f,I}(y))dy}W(f)\qquad x\in I'.
\end{equation}

Indeed it is easy to see that, if $f\in\cK_2(I)$, then
$\int_0^x(f(y)-p_{f,I}(y))dy$ does not depend on $x\in I'$, and is equal
to zero when $0\notin I$. Now let $0$ be in $I$, and suppose for
simplicity that $I\subset\Reali$.

Then formula (\ref{zeta}) may be
obtained by formula (\ref{theta}) using $R(\pi)$, the rotation by $\pi$,
which in the real line coordinates is $t\to-1/t$. As in
Proposition~\ref{dunitary}, such rotation is implemented by
$DU^{(2)}(R(\pi))D^*$ on the Hilbert space $\cH$, and, since $f$ has compact
support, we may set $x=\infty$ in formula (\ref{zeta}). Hence the
correspondence between (\ref{zeta}) and (\ref{theta}) follows by the
equality
 $$
2\int_0^\infty(f(y)-p_{f,I})(y)=
\langle\delta_0,2x\int_{0}^{-\frac{1}{x}} (f-p_{f,I})(y)dy
+f\left(-\frac{1}{x}\right)\rangle
$$

When $f$ is localized in $\Reali$ and we  chose $x=\infty$ as before,
the automorphisms $\zeta_c$ furnish extensions to the circle of the
restriction to $\cA_2(\Reali)$ of solitonic sectors on $\cA(\Reali)$. It is
not difficult to see that dilations act on these automorphisms dilating the
charge, therefore these sectors are non covariant too.

In the following subsection we shall see that the Weyl algebra on the
symplectic space $(X^2,\omega_2)$ is a global algebra for all these
sectors, i.e. the given automorphisms of the precosheaf modulo
unitaries on the Fock space are described by automorphisms of this global
algebra modulo inners. In doing that we shall see that the described sectors
form a group isomorphic to $\Reali^3$.

\subsection{Generalized BMT approach to sectors: Global description}
\label{subsec:globsect}

Now we describe some natural automorphisms of the Weyl algebra
$\cA^2(\dot\Reali)$ on $(X^2,\omega_2)$. If $\phi$ is a measure on $\dot\Reali$
such that $\int(1+t^2)d|\phi|(t)<\infty$, we set
 $$
\alpha^2_\phi(W(h))=e^{i\int h d\phi}W(h)\qquad h\in X^2,
 $$
and $\alpha^2_\phi$ extends to an automorphism of $\cA^2(\dot\Reali)$. This
automorphism is inner if and only if it is of the form ${\text{ad}}W(h)$ where
$h'''=\phi$, i.e. if and only if the first three moments (charges) of $\phi$
vanish:
 $$
Q_k(\phi)\doteq \int t^k d\phi(t)=0,\qquad k=0,1,2.
 $$
As a consequence, two such automorphisms are equivalent if all their charges
coincide. We now consider the corresponding sectors, i.e. classes of
automorphisms modulo inners.

\begin{Prop} The only covariant sector on $\cA^2(\Reali)$ in the above class is
the identity sector.
\end{Prop}

\begin{proof} First we see the behavior of the automorphisms under
translations:
\begin{align*}
Q_0(\beta_{T(t)}\alpha^2_\phi\beta_{T(-t)})
&=Q_0(\alpha^2_\phi)\cr
Q_1(\beta_{T(t)}\alpha^2_\phi\beta_{T(-t)})
&=\int(x-t)d\phi(x)=Q_1(\alpha^2_\phi)-t Q_0(\alpha^2_\phi)\cr
Q_2(\beta_{T(t)}\alpha^2_\phi\beta_{T(-t)})
&=\int(x-t)^2d\phi(x)=Q_2(\alpha^2_\phi)-2tQ_1(\alpha^2_\phi)+
t^2Q_0(\alpha^2_\phi).
\end{align*}
Then we compute the charges of automorphisms transformed with the ray inversion
$r:x\to-1/x$:
\begin{align*}
Q_0(\beta_{r}\alpha^2_\phi\beta_{r})
&=\int x^2d\phi(x)=Q_2(\alpha^2_\phi)\cr
Q_1(\beta_{r}\alpha^2_\phi\beta_{r})
&=\int x^2(-1/x)d\phi(x)=-Q_1(\alpha^2_\phi)\cr
Q_2(\beta_{r}\alpha^2_\phi\beta_{r})
&=\int x^2(-1/x)^2d\phi(x)=Q_0(\alpha^2_\phi).
\end{align*}
{}From the first equations we derive that a translation covariant sector has
$Q_0=Q_1=0$, while covariance under ray inversion amounts to $Q_0=Q_2$ and
$Q_1=0$, from which the thesis easily follows.
\end{proof}

\begin{rem}
If we generalize the preceding construction to the case of $n$
derivatives, thus obtaining sectors parameterized by $2n+1$ charges, the
preceding proof generalizes as well, then showing that the identity sector is
the only covariant sector in that case also.
\end{rem}

\begin{rem}
As BMT sectors, also the sectors described above are additive in the sense
that the vector charge $(Q_0,Q_1,Q_2)$ of the composition of two sectors
is just the
sum of the two charges. Then the action of $\PSL$ on these sectors gives a
linear representation of this group on $\Reali^3$. The absence of covariant
sectors means that the action is free and therefore has no
one-dimensional representations, i.e. it is irreducible.
\end{rem}

The local subalgebras for $\cA^2(\dot\Reali)$ are the
sub-algebras generated by the Weyl unitaries whose test functions are zero
outside $I$. Then the representation of $\cA^2(\Reali)$ on the Fock space
$e^{\cH^2}$ is faithful when restricted to local algebras, i.e. $\cA^2(I)$
may be
seen as a weakly dense subalgebra of the second quantization algebra of the
space $\cK^2(I)$.

By a classical Sobolev embedding argument, the functions in $\cK^2(I)$ are
continuous, therefore the automorphisms $\alpha^2_\phi|_{\cA^2(I)}$
uniquely extend to normal automorphisms of $\cR^2(I)$, so that $\alpha^2_\phi$
gives rise to an automorphism of the precosheaf $I\to\cR^2(I)$.

\begin{Prop}
All sectors described above may be localized in two points.
\end{Prop}
\begin{proof} First we observe that some of them may be localized even in one
point, in fact the multiples of the $\delta_0$ function
give sectors with $Q_1=Q_2=0$, while for the measures $cx^{-2}\delta_\infty(x)$
we have $Q_0=0$, $Q_1=0$, $Q_2=c$, therefore we may restrict to the case
$(Q_0,Q_1)\not=(0,0)$.
Then we have to show that for any triple $Q_i$, $i=0,1,2$,
$(Q_0,Q_1)\not=(0,0)$, we may find a measure $\phi=\lambda\delta_a+\mu\delta_b$
with the given momenta for some $\lambda,\mu,a,b\in\Reali$, or equivalently
solve the system
 $$
\left\{
\begin{matrix}
Q_0&=\lambda+\mu\cr
Q_1&=\lambda a+\mu b\cr
Q_2&=\lambda a^2+\mu b^2\cr
\end{matrix}\right.\ ,
 $$
whose solutions are obtained choosing a $b$ for which $Q_1-Q_0b\not=0$ and
$Q_0b^2-2Q_1b+Q_2\not=0$ and then setting
$a=(Q_2-Q_1b)(Q_1-Q_0b)^{-1}$, $\lambda=(Q_1-Q_0b)^2(Q_0b^2-2Q_1b+Q_2)^{-1}$,
$\mu=Q_0-\lambda$.
\end{proof}

The preceding proposition constitutes indeed another proof that these sectors
are non covariant, since the following theorem holds:

\begin{Prop}Let $I\to\cA(I)$ a local conformal precosheaf on $S^1$. Then a
covariant sector $\rho$ with finite index which may be localized in two points
is trivial.
\end{Prop}

\begin{proof} We may suppose that the two points are $\{0,\infty\}$.
We first observe that $\rho$ is indeed an automorphism because, if $j$ is the
antiunitary modular conjugation for $\cR(0,\infty)$, $j\rho j$ is still
localized in the same two points and then any intertwiner between $\rho j\rho j$
and the identity (which exists e.g. by \cite{[GuLo2]}) is localized in
these two points
and is therefore a number by two-regularity.

The same argument shows that $\rho$ commutes with the dilations, because the
cocycle in the covariance equation for the dilation group is then trivial.

Now the state $\omega_0\cdot\rho^{-1}$ is dilation invariant, and therefore, by
a cluster argument, coincides with the vacuum state, which ends the proof.
\end{proof}

In this last part of the subsection we show the relation between the local and
the global picture of the superselection sectors of the first-derivative theory
or, more precisely, we show that all the sectors described in subsection
\ref{subsec:locsect} are (normal extensions of) the sectors of
$\cA^2(\dot\Reali)$ described here.

\begin{Prop} The sectors $[\alpha_{\phi,\lambda}]$, $[\theta_c]$
$[\zeta_c]$ are of the form $[\alpha^2_\mu]$, $\mu$ measure.
\end{Prop}

\begin{proof} Given a non trivial sector $[\alpha_{\phi,\lambda}]$, we may
choose a representative s.t. supp~$\phi\subset I_0$, where $I_0$ is a given
compact interval in $\Reali$. In order $[\alpha_{\phi,\lambda}]$ to be localized
in $I_0$, we should have $\langle\phi,f\rangle_{I_0,\lambda}=0$ for any $f$
localized in $I'_0$, i.e., since on $I'_0$ $f$ coincides with $p_{f,I_0}$, we
get $\int\phi(x)(x-\lambda)dx=0$, i.e. $\lambda=\lambda_0=\frac{\int
x\phi(x)}{\int\phi(x)}$ (the denominator does not vanish since
$[\alpha_{\phi,\lambda}]$ is non trivial). Then, according to formula
(\ref{difference}), one has
 $$
[\alpha_{\phi,\lambda}]=
[\theta_{-Q(\lambda-\lambda_0)}]\cdot [\alpha_{\phi,\lambda}]
 $$
hence, since the class $\{\alpha_\mu$, $\mu$ measure$\}$ is closed under
composition, it is enough to prove the statement for
$[\alpha_{\phi,\lambda_0}]$, $[\theta_c]$ and $[\zeta_c]$.

As far as $\alpha_{\phi,\lambda_0}$ is concerned,
we have $\int\phi(x)p_{f,I}(\lambda_0)dx$ $=$ $\int\phi(x)p_{f,I}(x)dx$,
therefore $\langle\phi,f\rangle_{I,\lambda_0}$
coincides with $\int\phi(x)(f(x)-p_{f,I}(x))dx$ and therefore, integrating
by
parts, with $-\int(\int(f-p_{f,I}))d\phi(x)$. Observing that
$\int(f-p_{f,I})$ is exactly the representative in $X^2$ of
$D^{-1}[f]$
which vanishes outside $I$ we conclude that the automorphism
$\alpha_{\phi,\lambda_0}$ comes from the automorphism $\alpha^2_{-d\phi}$ on
$\cA^2(\dot\Reali)$, and the relation among the charges is
$Q_0(\alpha^2_{-d\phi})=0$,
$Q_1(\alpha^2_{-d\phi})=Q(\alpha_{\phi,\lambda})$,
$Q_2(\alpha^2_{-d\phi})=2\lambda_0 Q(\alpha_{\phi,\lambda})$.

In the same way we may show that the ``solitonic'' automorphisms
$\zeta_c$ in equation~(\ref{zeta}) come from the automorphisms
$\alpha^2_{\mu}$ on $\cA^2(\Reali)$ with $\mu=-c\delta_0.$
Conjugating $\zeta_c$ with the ray inversion we see that
 the automorphisms $\theta_c$ in
equation~(\ref{theta}) localized at
infinity come from the automorphisms $\alpha^2_\mu$ on $\cA^2(\dot\Reali)$
with $\mu=cx^{-2}\delta_\infty$.
\end{proof}

\begin{rem}  In \cite{[DF]}
it is shown that in
any diffeomorphism covariant theory  on $S^1$, in physics terms
a theory with a  stress-energy tensor, superselection sectors are
covariant.  It is well known that the usual way to associate a
stress-energy tensor to the derivative of the $U(1)$-current formally leads
to a
conformal charge $c=\infty$. Our result then shows
that the $U(1)$-current  derivative theories indeed
do not have a stress energy tensor.
\end{rem}
\medskip

{\bf Acknowledgment.}
H.-W. W. wishes to thank the University of Rome II for the kind hospitality
and the CNR for financial support during a visit in Rome where this
collaboration started. He also wants to thank Bert Schroer warmly
for various helpful discussions.

\end{document}